\pdfoutput=1 
\documentclass{JINST}
\let\ifpdf\relax
\usepackage{graphics}
\usepackage{epsfig}

\usepackage{amssymb}

\usepackage{wrapfig}
\usepackage{capt-of}

\usepackage{caption}
\usepackage{subcaption}
\usepackage{float}

\usepackage{color}
\definecolor{pink}{rgb}{1.,0.75,0.8}
\definecolor{green}{rgb}{0.3,1,0.3}
\definecolor{dgreen}{rgb}{0.,0.6,0.}
\definecolor{gold}{rgb}{1.,0.84,0.}
\definecolor{beige}{rgb}{0.96,0.96,0.86}
\definecolor{myyellow}{rgb}{1.,0.84,0.8}
\definecolor{grey}{rgb}{0.8,0.8,0.8}
\definecolor{darkyellow}{cmyk}{0,0,0.5,0.5}
\definecolor{darkwhite}{gray}{0.9}
\definecolor{lightblack}{gray}{0.1}
\definecolor{black}{gray}{0}
\definecolor{lightblue}{rgb}{0.1,0.4,1.0}
\definecolor{fucsia}{rgb}{1.,0.4,0.9}
\definecolor{red}{rgb}{1.,0.,0.}

\title{Conceptual design and simulation of a water Cherenkov muon veto for the XENON1T experiment}

\author{E.~Aprile$^a$,
F.~Agostini$^{e,b}$,
M.~Alfonsi$^c$,
K.~Arisaka$^d$,
F.~Arneodo$^e$\footnote{Present address: New York University in Abu Dhabi, UAE},
M.~Auger$^f$,
C.~Balan$^g$,
P.~Barrow$^f$,
L.~Baudis$^f$,
B.~Bauermeister$^h$,
A.~Behrens$^f$,
P.~Beltrame$^i$\footnote{Present address: School of Physics \& Astronomy, The University of Edinburgh, Edinburgh, United Kingdom},
K.~Bokeloh$^l$,
A.~Breskin$^i$,
A.~Brown$^m$,
E.~Brown$^l$,
S.~Bruenner$^n$,
G.~Bruno$^e$,
R.~Budnik$^i$,
J.~M.~R.~Cardoso$^g$,
A.P.~Colijn$^c$,
H.~Contreras$^a$,
J.P.~Cussonneau$^o$,
M.P.~Decowski$^c$,
E.~Duchovni$^i$,
S.~Fattori$^h$\thanks{Corresponding author.}~,
A.~D.~Ferella$^e$,
W.~Fulgione$^p$,
M.~Garbini$^b$,
C.~Geis$^h$,
L.~W.~Goetzke$^a$,
C.~Grignon$^h$,
E.~Gross$^i$,
W.~Hampel$^n$,
R.~Itay$^i$,
F.~Kaether$^n$,
G.~Kessler$^f$,
A.~Kish$^f$,
H.~Landsman$^i$,
R.~F.~Lang$^m$,
M.~Le~Calloch$^o$,
D.~Lellouch$^i$,
L.~Levinson$^i$,
C.~Levy$^l$,
S.~Lindemann$^n$,
M.~Lindner$^n$,
J.~A.~M.~Lopes$^g$\footnote{also with Coimbra Engineering Institute, 3030-199 Coimbra, Portugal},
K.~Lung$^d$,
A.~Lyashenko$^d$,
S.~MacMullin$^m$,
T.~Marrod\'an~Undagoitia$^n$,
J.~Masbou$^o$,
F.~V.~Massoli$^b$,
D.~Mayani~Paras$^f$,
A.~J.~Melgarejo~Fernandez$^a$,
Y.~Meng$^d$,
M.~Messina$^a$,
B.~Miguez$^p$,
A.~Molinario$^{e,p}$,
G.~Morana$^b$,
M.~Murra$^l$,
J.~Naganoma$^q$,
U.~Oberlack$^h$,
S.~E.~A.~Orrigo$^g$\footnote{Present address: IFIC, CSIC-Universidad de Valencia, Valencia, Spain},
E.~Pantic$^d$\footnote{Present address: Department of Physics, UC Davis, Davis, CA , USA},
R.~Persiani$^b$,
F.~Piastra$^f$,
J.~Pienaar$^m$,
G.~Plante$^a$,
N.~Priel$^i$,
S.~Reichard$^m$,
C.~Reuter$^m$,
A.~Rizzo$^a$,
S.~Rosendahl$^l$,
J.~M.~F.~dos~Santos$^g$,
G.~Sartorelli$^b$,
S.~Schindler$^h$,
J.~Schreiner$^n$,
M.~Schumann$^r$,
L.~Scotto~Lavina$^o$,
M.~Selvi$^b$,
P.~Shagin$^q$,
H.~Simgen$^n$,
A.~Teymourian$^d$,
D.~Thers$^o$,
A.~Tiseni$^c$,
G.~Trinchero$^p$,
O.~Vitells$^i$,
H.~Wang$^d$,
M.~Weber$^n$,
C.~Weinheimer$^l$.\\
(The XENON Collaboration)\\
\llap{$^a$}Physics Department, Columbia University, New York, NY, USA \\
\llap{$^b$}Department of Physics, University of Bologna and INFN-Bologna, Bologna, Italy \\
\llap{$^c$}Nikhef and the University of Amsterdam, Science Park, Amsterdam, Netherlands \\
\llap{$^d$}Physics \& Astronomy Department, University of California,  Los Angeles, CA, USA \\
\llap{$^e$}INFN-Laboratori Nazionali del Gran Sasso and Gran Sasso Science Institute, 67100 L'Aquila, Italy \\
\llap{$^f$}Physik-Institut, University of Zurich, Zurich, Switzerland \\
\llap{$^g$}Department of Physics, University of Coimbra, Coimbra, Portugal \\
\llap{$^h$}Institut f\"ur Physik \& Exzellenzcluster PRISMA, Johannes Gutenberg-Universit\"at Mainz, Mainz, Germany \\
\llap{$^i$}Department of Particle Physics and Astrophysics, Weizmann Institute of Science, Rehovot, Israel \\
\llap{$^l$}Institut f\"ur Kernphysik, Wilhelms-Universit\"at M\"unster, M\"unster, Germany \\
\llap{$^m$}Department of Physics and Astronomy, Purdue University, West Lafayette, IN, USA \\
\llap{$^n$}Max-Planck-Institut f\"ur Kernphysik, Heidelberg, Germany \\
\llap{$^o$}SUBATECH, Ecole des Mines de Nantes, CNRS/In2p3, Universit\'e de Nantes, Nantes, France \\
\llap{$^p$}INFN-Torino and Osservatorio Astrofisico di Torino, Torino, Italy \\
\llap{$^q$}Department of Physics and Astronomy, Rice University, Houston, TX, USA \\
\llap{$^r$}Albert Einstein Center for Fundamental Physics, University of Bern, Bern, Switzerland \\
E-mail: \email{dr.serena.fattori@gmail.com}}

\abstract{XENON is a dark matter direct detection project, consisting of a time projection chamber (TPC) filled with liquid xenon as detection medium. The construction of the next generation detector, XENON1T, is presently taking place at the Laboratori Nazionali del Gran Sasso (LNGS) in Italy. It aims at a sensitivity to spin-independent cross sections of $2 \cdot 10^{-47} ~ \mathrm{cm}^{\mathrm{2}}$ for WIMP masses around 50~GeV/c$^{2}$, which requires a background reduction by two orders of magnitude compared to XENON100, the current generation detector. An active system that is able to tag muons and muon-induced backgrounds is critical for this goal. A water Cherenkov detector of $\sim$10 m height and diameter has been therefore developed, equipped with 8~inch photomultipliers and cladded by a reflective foil. We present the design and optimization study for this detector, which has been carried out with a series of Monte Carlo simulations. The muon veto will reach very high detection efficiencies for muons ($>99.5\%$) and showers of secondary particles from muon interactions in the rock ($>70\%$). Similar efficiencies will be obtained for XENONnT, the upgrade of XENON1T, which will later improve the WIMP sensitivity by another order of magnitude. With the Cherenkov water shield studied here, the background from muon-induced neutrons in XENON1T is negligible.}


\begin{document}

\section{Introduction} \label{sec:introduction}
The XENON project is devoted to the direct detection of Weakly Interactive Massive Particles (WIMPs), a prime cold dark matter candidate arising, e.g., from theories of supersymmetry (SUSY) or Universal Extra Dimensions (UED) \cite{Feng}.

Recently the results from noble liquid detectors have shown the promising performances of this technology. The first experiment of the XENON project, based on the prototype detector XENON10 operated at LNGS \cite{XENON10_SI, XENON10_SD}.
The subsequent detector, XENON100 (of 100 kg scale mass), is currently taking dark matter data at LNGS. XENON100  completed a dark matter run with 225 live~days of data  in 2012, leading to a limit for spin-independent elastic WIMP-nucleon scattering cross~section of $2.0 \cdot 10^{-45} ~\mathrm{cm}^{\mathrm {2}}$ for a WIMP mass of 55~GeV/c$^{\mathrm{2}}$ at 90\% C.L. \cite{XENON100_225days_SI} and, in 2013, for spin-dependent WIMP-neutron scattering cross~section of $3.5 \cdot 10^{-40} ~ \mathrm{cm}^{\mathrm{2}}$ for a WIMP mass of 45~GeV/c$^{\mathrm{2}}$ at 90\% C.L. \cite{XENON100_255days_SD}. 

During the operation of XENON100, the design and construction of XENON1T is taking place. This new generation detector aims to reach a sensitivity goal of $2 \cdot 10^{-47} ~ \mathrm{cm}^{\mathrm{2}}$ for a WIMP mass of 50~GeV/c$^{\mathrm{2}}$ at 90\% C.L. \cite{DM2012_Ranny}. To achieve this goal, a background reduction of two orders of magnitude compared to XENON100 \cite{em_bk_xe100, nr_bk_xe100}  is required.

\begin{figure}[h]
\centering
\includegraphics[height=5.6cm] {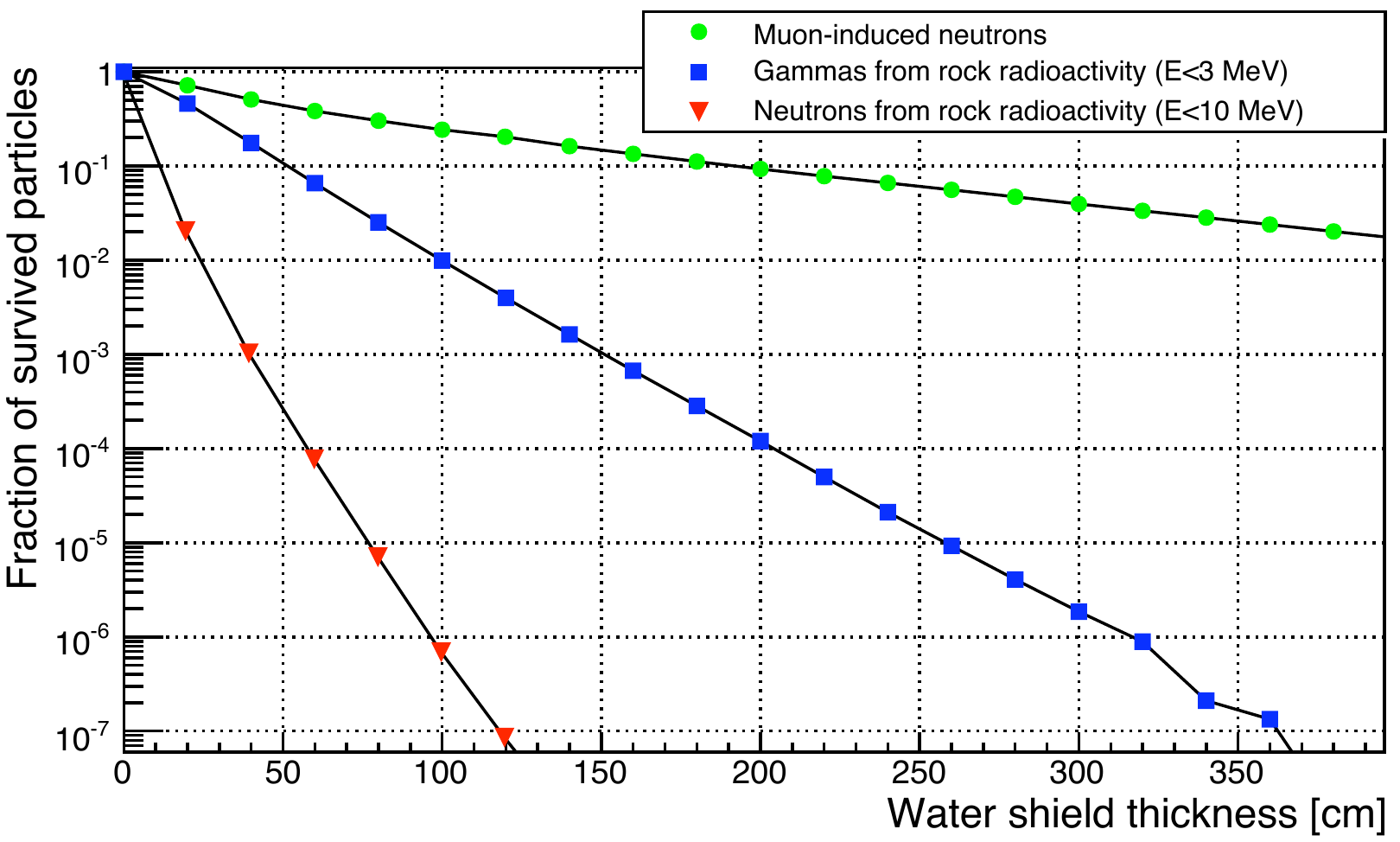}
\caption{Fraction of survived particles as a function of the thickness of the water shield surrounding the detector. Circular dots for the muon-induced neutrons, squared dots for the gammas from rock radioactivity and triangular dots for neutrons from rock radioactivity.}
\label{fig:shield}
\end{figure}

Next to the design of the detector and the selection of radiopure materials, the shielding against cosmic rays and natural radioactivity are critical to obtain the sensitivity.
The XENON1T detector is being placed underground in Hall B of LNGS.
The residual muon flux reaching the experimental hall is $(3.31 \pm 0.03) \cdot 10^{-8} ~ \mu / (\mathrm{cm}^{\mathrm{2}}$ $\mathrm{s})$ (according to \cite{Aglietta, VitalyLuciano, ICRCSelvi}) with an average energy of $\sim$270~GeV \cite{MACRO}.
Moreover, the detector will be installed in the center of a large cylindrical tank filled  with water. Figure \ref{fig:shield} is the result of a  simulation showing the absorption of neutrons and $\gamma$ from rock radioactivity and of muon-induced neutrons as a function of the water shield thickness. A few meters of water constitute an effective shield against gammas and neutrons produced by rock radioactivity. The only residual background after 4 meters of water is given by muon-induced neutrons, which are produced via direct muon spallation of nuclei or through electromagnetic and hadronic cascades generated by muons.
A conservative estimate of the muon-induced neutron flux in the LNGS cavern is  $\sim 7.3 \cdot 10^{-10} ~ \mathrm{n} / (\mathrm{cm}^{\mathrm{2}} $ $\mathrm{s})$   for $\mathrm{E_{n}} > 10 ~ \mathrm{ MeV}$ \cite{MeiHime}, that is about 3 orders of magnitude lower than that of neutrons from concrete radioactivity, but their energy spectrum extends up to tens of GeV. They may constitute a dangerous background since they can cross the water shield and scatter elastically off the target nuclei leaving a WIMP-like signal.  
This fact motivated instrumenting the water tank (WT) as an active muon-veto to detect the Cherenkov light that charged particles produce as they cross the water. The aim is to identify events in which a muon directly crosses the WT, and also events in which a muon is outside but the particles constituting the electromagnetic or hadronic cascade enter the WT.

In this work we present the study performed to design the muon-veto for the XENON1T experiment. A detailed Monte Carlo simulation was set up in order to optimize the working configuration. In particular, the total number and the arrangement of photomultipliers (PMTs) and the possibility to use a reflective foil were investigated. Following the outcomes of the simulation, different possible trigger conditions were evaluated.
The study was extended to the case of a future upgrade of XENON1T, the XENONnT experiment, that will aim to an improvement in sensitivity of one order of magnitude by doubling the xenon mass.

\section{Muon veto conceptual design} \label{sec:dete}
The ``ingredients'' of the XENON1T water Cherenkov muon veto are a WT, $\sim$~10~m heigh and  $\sim$~10~m in diameter, a certain number of PMTs, with a certain geometrical arrangement and working with an appropriate trigger condition, and a reflective foil. 

\begin{figure}[t!]
\hspace{-0.25cm}
	\begin{minipage}[t]{0.49\linewidth} 
		\centering
			   \includegraphics[height=6.1cm]{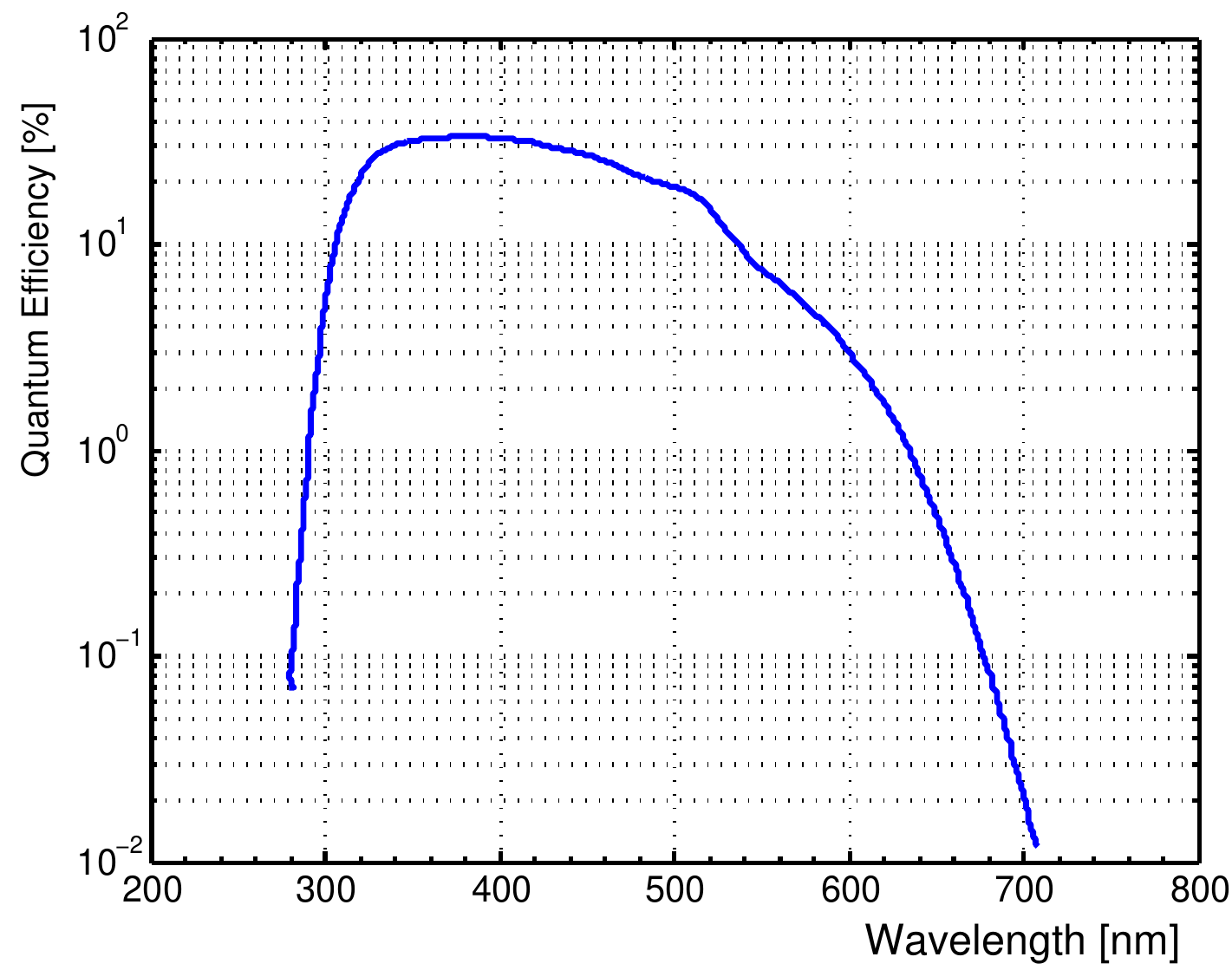} 
  \caption{Quantum efficiency of one of the HQE Hamamatsu PMTs R5912ASSY \cite{Hamamatsu}.} 
  \label{fig:QE_PMTs} 
	\end{minipage}
		\hspace{0.20cm} 
	\begin{minipage}[t]{0.49\linewidth}
	\centering
	  \includegraphics[height=6.0cm]{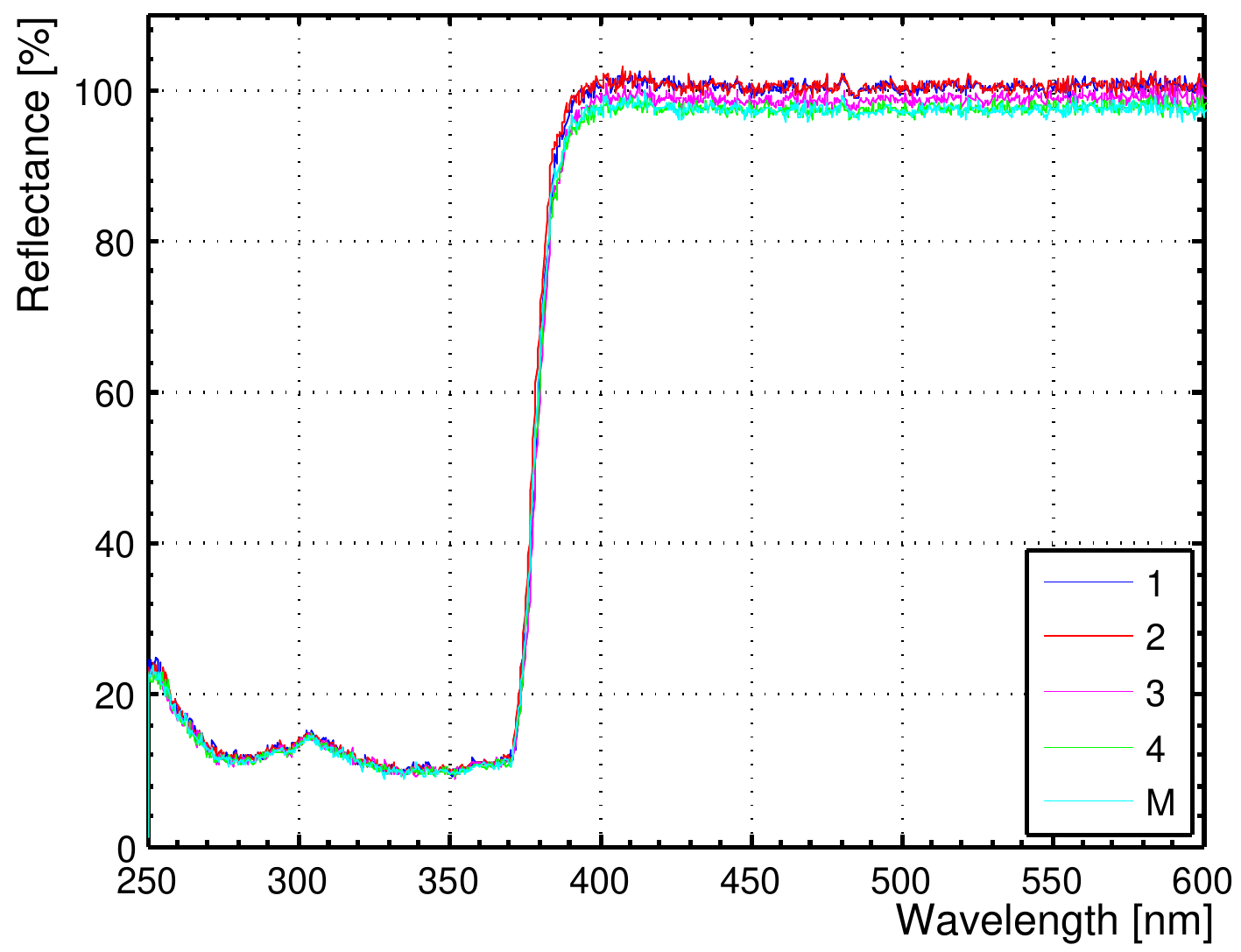} 
  \caption{\hspace{0.20cm} Specular reflectivity measurements of DF2000MA in different positions on the surface of a square shaped foil sample (Edges: 1-4, Middle: M) \cite{Chris}.} 
  \label{reflectivity} 
	\end{minipage}
\end{figure}

The selected PMT is the high quantum efficiency (HQE) 8''~Hamamatsu  R5912ASSY, already provided with a water-proof enclosure. These PMTs have a bialkali photocathode and borosilicate glass window. Ten dynodes provide a typical gain of $10^{7}$ at a working voltage of {$\sim$1500~V}. The quantum efficiency is about 30\%, averaged over the Cherenkov light wavelength distribution, in the range {[300-600]~nm} (see figure~\ref{fig:QE_PMTs}), and the collection efficiency is 85\%, as declared by the manufacturer  \cite{Hamamatsu}.  

The reflective foil chosen is DF2000MA by 3M, which provides a very good reflectivity (more than 99\% between $\sim$400 and $\sim$1000~nm, as declared by the manufacturer \cite{3M} and consistent with what obtained with dedicated experimental measurements, as shown in figure~\ref{reflectivity}), and allows for a shift in the wavelength of the UV Cherenkov photons toward the blue region in order to better match with the PMTs' wavelength sensitivity. A similar study was carried out in the GERDA experiment, using the same reflective foil by 3M (at the time called ``VM2000'') \cite{GERDA}.

\vspace{1.0cm}

\section{Description of the Monte Carlo simulation} \label{sec:MC}
\subsection{Physics inputs}
The optimization of the muon veto design has been obtained with a study done through a detailed Monte Carlo simulation based on GEANT4 \cite{GEANT4} version 4.9.3.

Our GEANT4 Physics List considered the following processes:
\begin{itemize}
	\item muon-induced spallation (or muon photo-nuclear interaction) modeled above 1~GeV muon energy; the final-state generator relies on parametrized hadronic models;
	\item gamma inelastic scattering (the photo-nuclear interaction) generating its hadronic final states through a chiral-invariant  phase-space decay model below 3~GeV; a theoretical quark-gluon string model simulates the punch-through reactions at higher energies;
	\item hadronic interactions of nucleons, pions and kaons simulated with the quark-gluon string model above 6~GeV, an intra-nuclear binary cascade model at lower energies and a pre-equilibrium de-exitation stage below 70~MeV;
	\item the equilibrium stage considering fragment and gamma evaporation, fission, Fermi break-up and multi-fragmentation of highly-excited nuclei;
	\item neutron transport and interaction described by data-driven models below 20~MeV: High-Precision (HP) model;
	\item elastic scattering of hadrons above 20~MeV described by the \textit{G4LElastic} model;
	\item standard treatment of the electromagnetic processes.
\end{itemize}

\begin{figure}[!b] 
  \centering 
   \includegraphics[height=6.0cm]{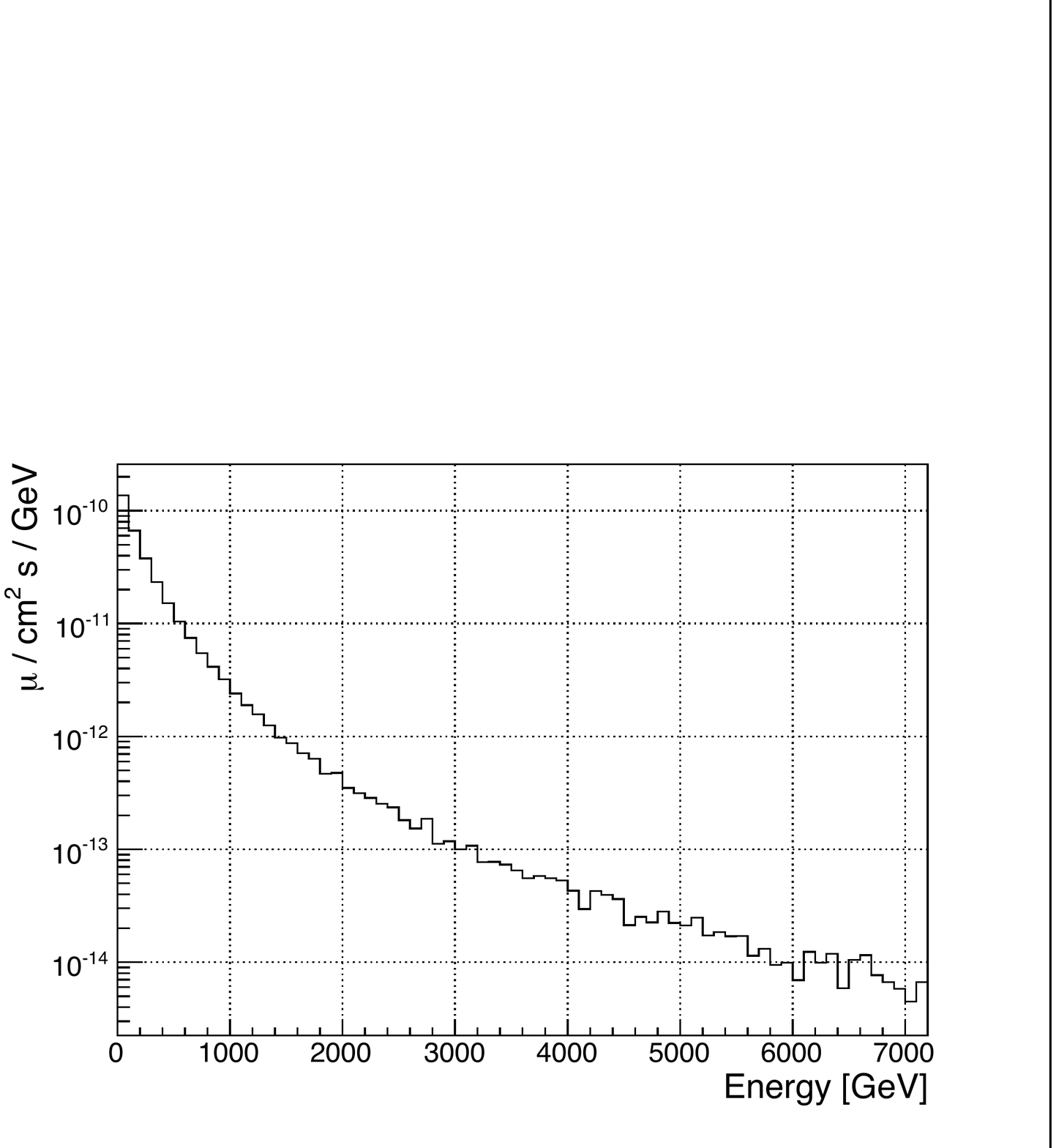}  
  \caption{Muon energy distribution at LNGS underground laboratory \cite{Aglietta2}.} 
  \label{mc_FirstStep_energy} 
\end{figure} 

\begin{figure}[!t]
	\centering
	\begin{tabular}{cc} 

		\hspace{-0.7cm}
  		\begin{tabular}{c}
  			   \includegraphics[width=0.48\textwidth]{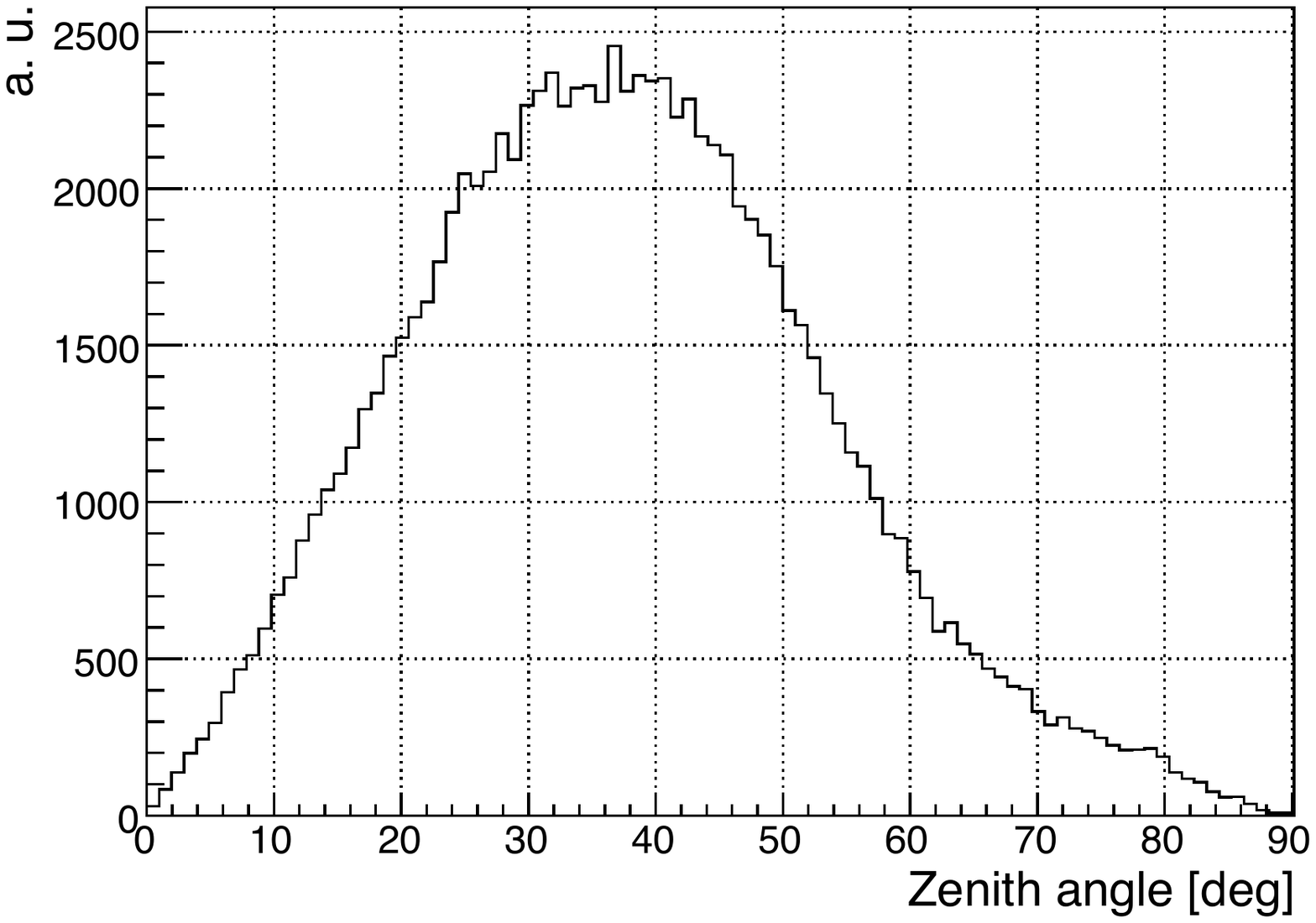}  
		\end{tabular}  

		&

		\hspace{0.3cm}  
  		\begin{tabular}{c}
  			\includegraphics[width=0.48\textwidth]{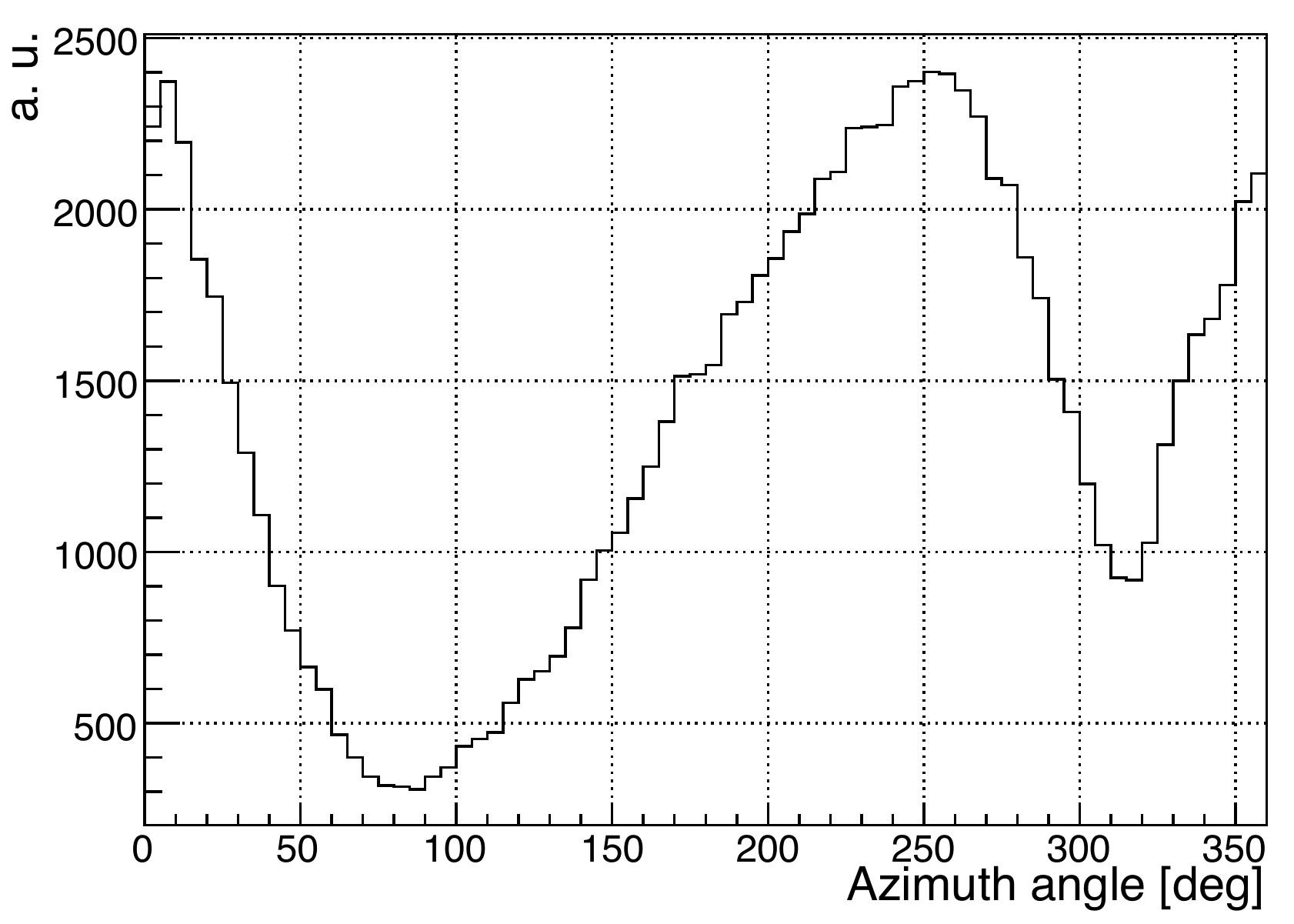} 
		\end{tabular}

		\\

		\hspace{-0.4cm}
		\scriptsize{(a) Zenith angle}

		 &

		\scriptsize{(b) Azimuth angle}
		
		\\
	
	\end{tabular}
  \caption{Muon angular distributions at LNGS underground laboratory \cite{Aglietta2}.}
   \label{mc_FirstStep_angular}
\end{figure}

As a first step 2~$ \cdot 10^{8}$ muons were generated, using the energy and angular distribution shown in figures  \ref{mc_FirstStep_energy} and \ref{mc_FirstStep_angular}. The muons were sampled over a large area, a circle with 30~m radius, and were propagated through at least 5~m of rock around the experimental hall in order to guarantee the full development of the muon-induced showers. With the measured muon flux at LNGS, this number of simulated events over the sampling area corresponds to about 7~years. Our MC simulation predicts a neutron flux in the cavern of  1.1~$\cdot$ 10$^{-10}$~n/(cm$^{2}$~s) for E$_{n}>$~10~MeV. However, since it is known that the muon-induced neutron production in rock is underestimated in GEANT4 \cite{MeiHime}, we conservatively scaled up our results to the highest value found in literature \cite{MeiHime}, namely 7.3~$\cdot10^{-10}$~n/(cm$^{2}$~s), a factor 7 larger. 

Among all the events with at least one neutron hitting the WT, one third of the neutrons come from events where also the muon crosses the WT (hereafter called "\textit{muon events}"), while the remaining ones are either neutrons alone, or accompanied by an electromagnetic or hadronic shower ("\textit{shower events}").

For the second step, 10$^{4}$ ``muon events'' and 10$^{4}$ ``shower events'' were injected into a simulation containing a detailed description of the XENON1T geometry. It features the water shield and all detector elements such as the cryostat and the cryostat support structure, which could shadow the photons from the Cherenkov light generated in the water. At the surface of the WT, each event was generally composed of many particles, with at least one muon-induced neutron. 

In particular the ``shower event'' case is more challenging, because the muon-induced neutron has to be tagged
with the much more faint Cherenkov light produced by secondary particles, compared to the muon
signal.

\vspace{0.5cm}

\subsection{Muon veto configurations}
We studied several detector configurations, varying the type of reflection and the reflectivity percentage of the foil, the number of PMTs and their geometric arrangement, all constrained by budget considerations. 

\begin{figure}[H]
\centering
	\begin{minipage}[t]{0.3\linewidth} 
		\centering
		\includegraphics[width= 0.95\textwidth]{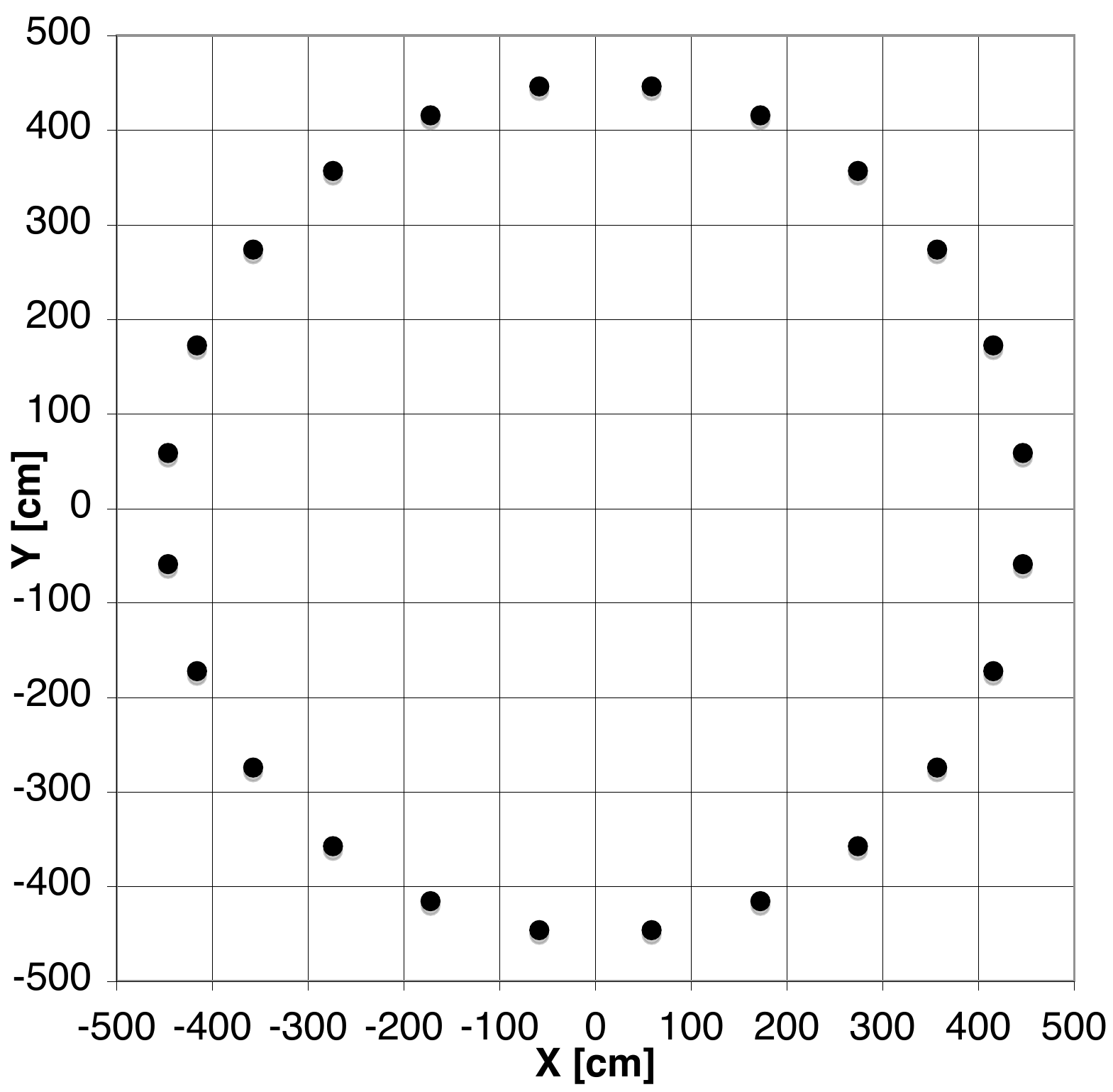}\\
		\footnotesize{\textsf{\textbf{a.} Bottom / Top Array Model 1 with 24 PMTs.}}
	\end{minipage}
	\hspace{0.5cm} 
	\begin{minipage}[t]{0.3\linewidth} 
		\centering
		\includegraphics[width= 0.95\textwidth]{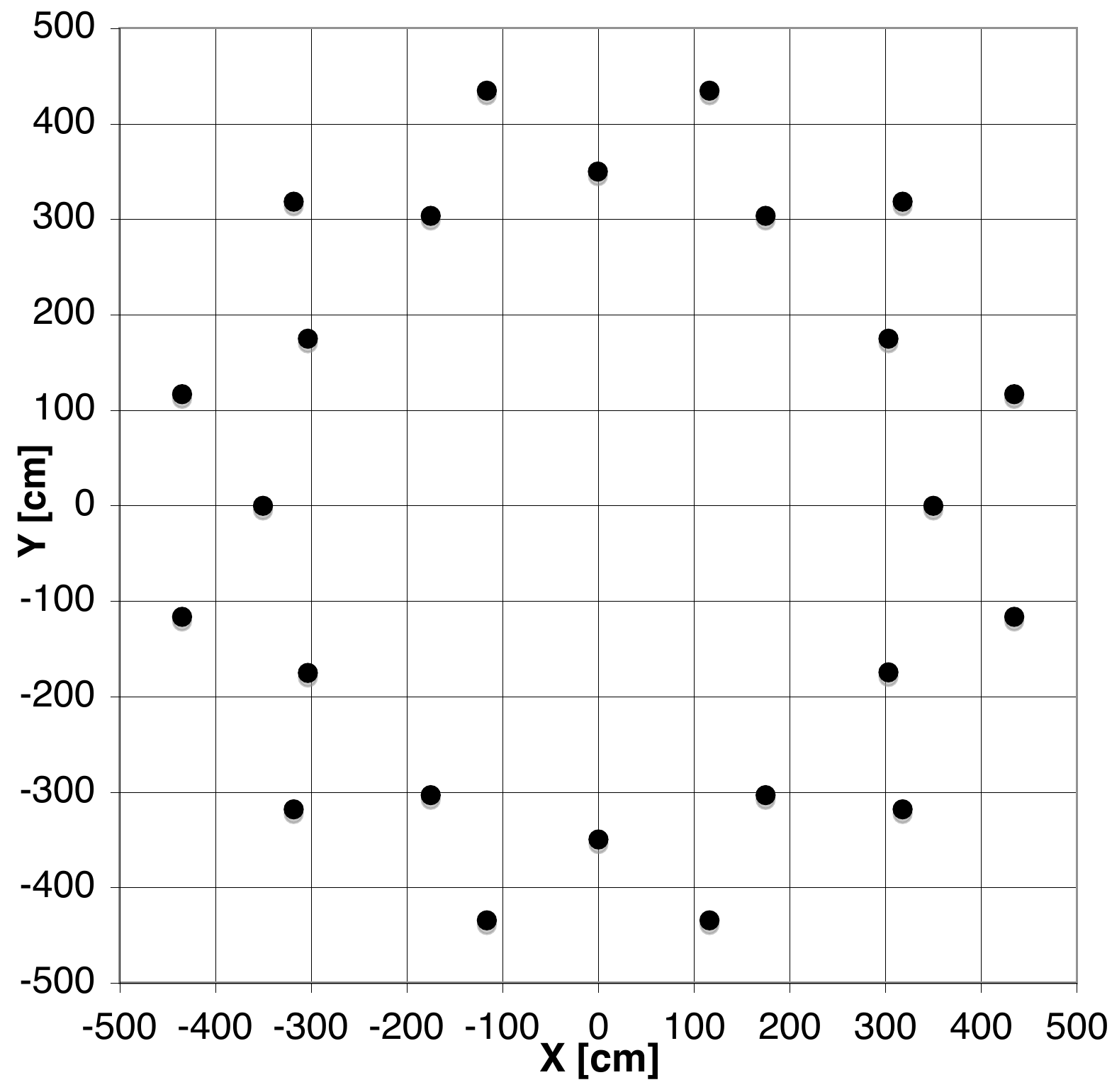}\\
		\footnotesize{\textsf{\textbf{b.} Bottom / Top Array Model 2 with 24 PMTs.}}
	\end{minipage}
	\hspace{0.5cm} 
	\begin{minipage}[t]{0.3\linewidth} 
		\centering
		\includegraphics[width= 0.95\textwidth]{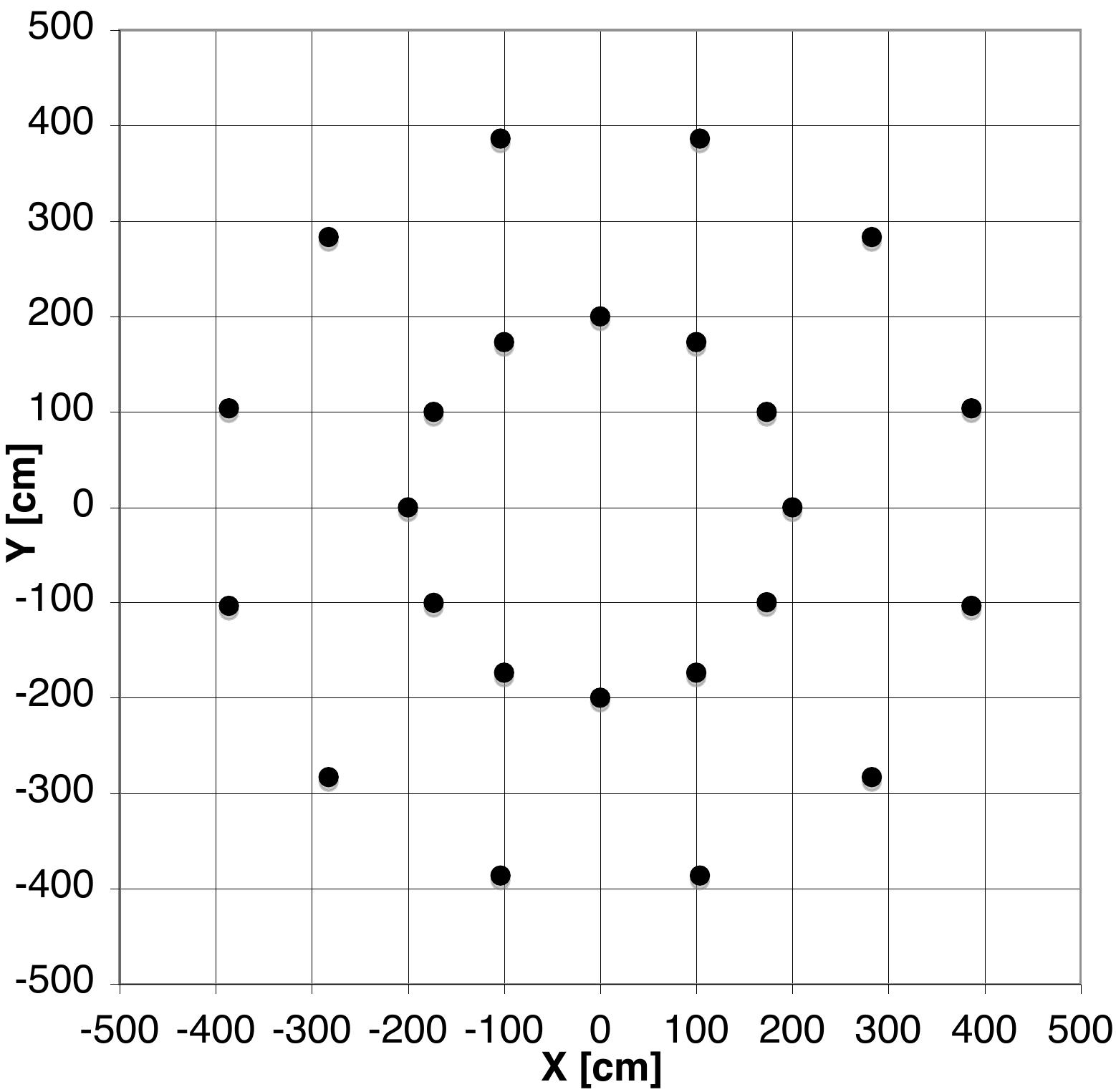}\\
		\footnotesize{\textsf{\textbf{c.} Bottom / Top Array Model 3 with 24 PMTs.}}
	\end{minipage}\\
	\vspace{0.5cm}
	\begin{minipage}[t]{0.3\linewidth} 
		\centering
		\includegraphics[width= 0.95\textwidth]{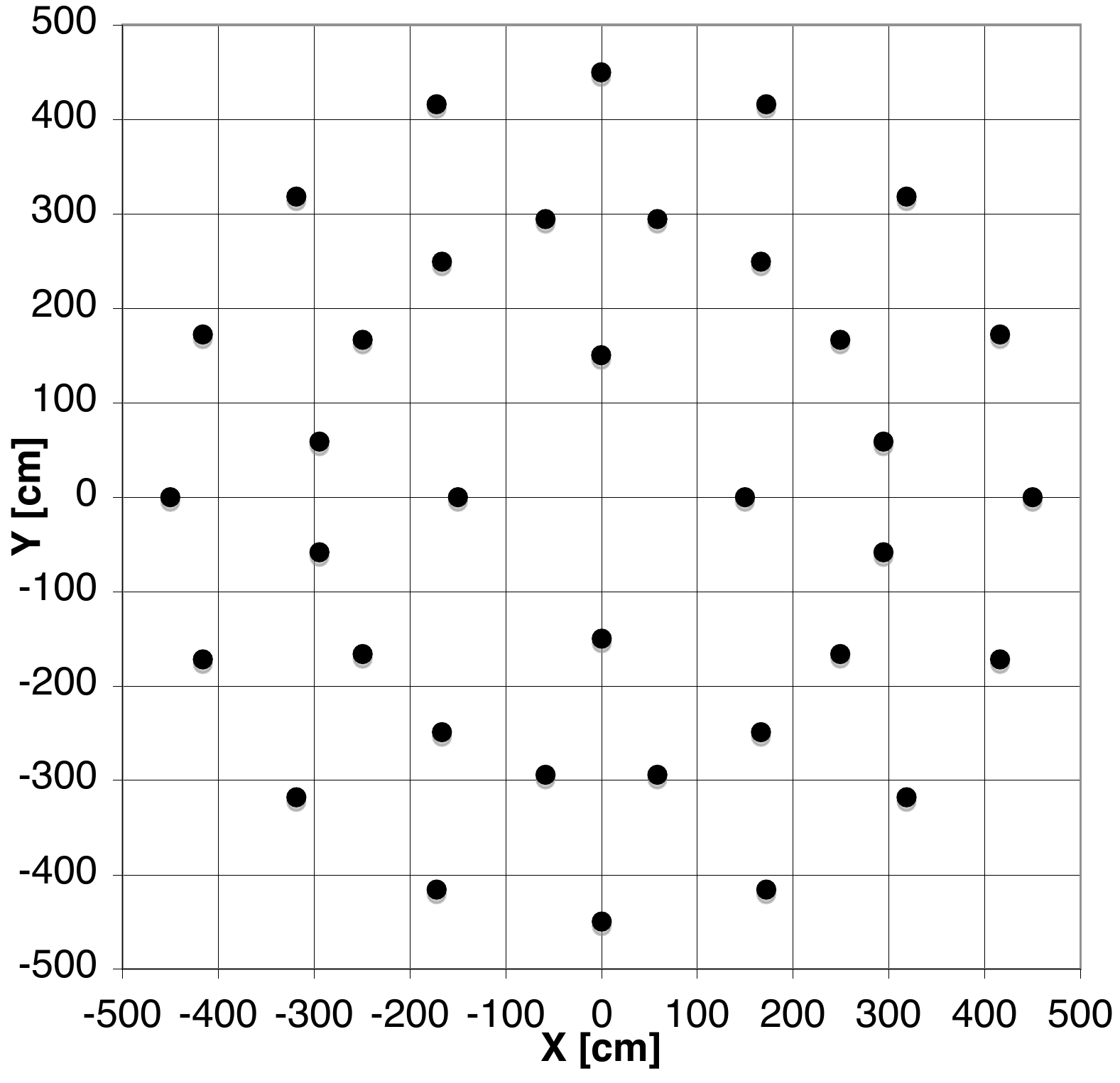}\\
		\footnotesize{\textsf{\textbf{d.} Bottom / Top Array Model 4 with 36 PMTs.}}
	\end{minipage}
	\hspace{0.5cm} 
	\begin{minipage}[t]{0.3\linewidth} 
		\centering
		\includegraphics[width= 0.95\textwidth]{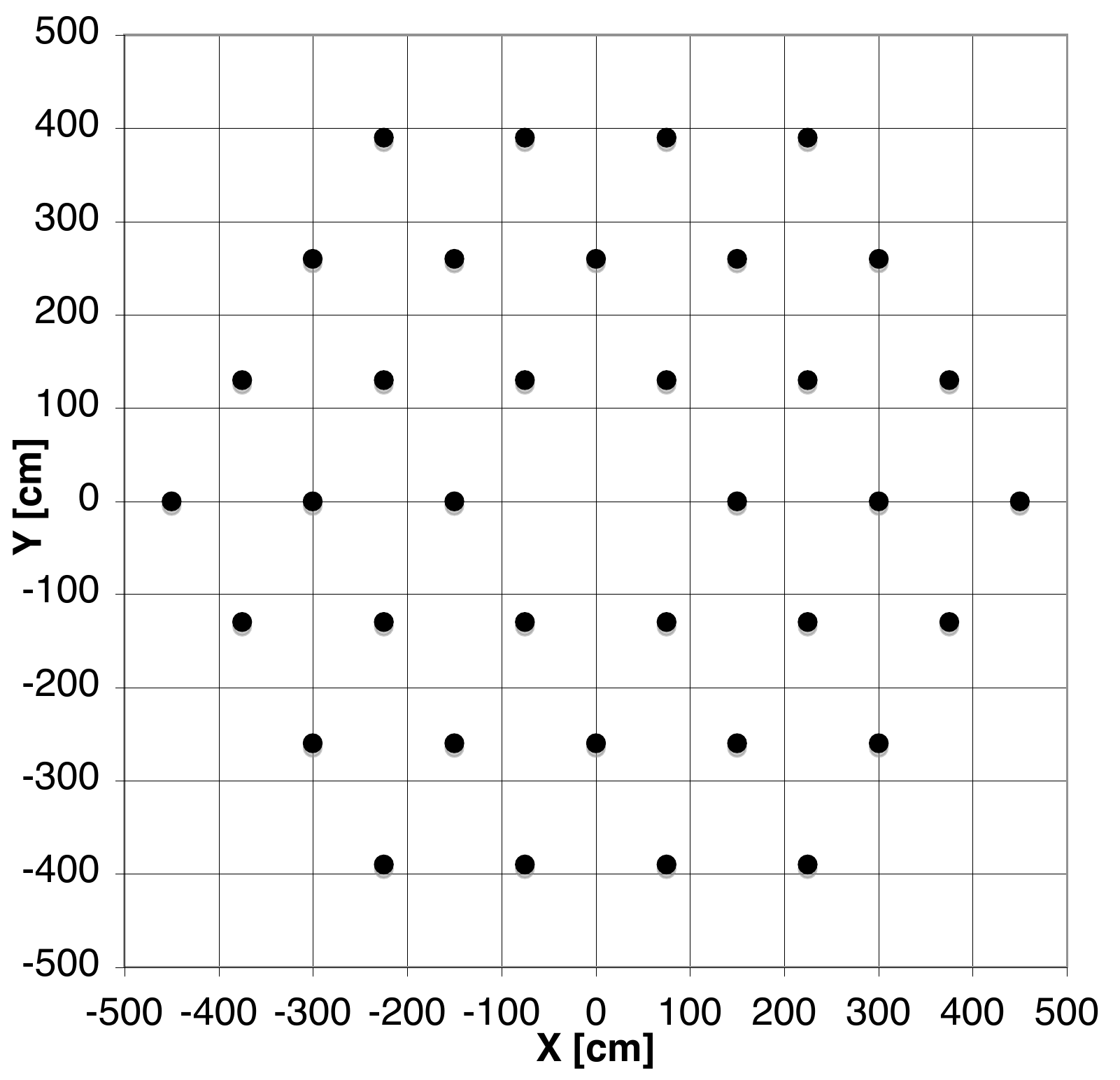}\\
		\footnotesize{\textsf{\textbf{e.} Bottom / Top Array Model 5 with 36 PMTs.}}
	\end{minipage}\\
	\vspace{0.5cm}
	\begin{minipage}[t]{1.0\linewidth} 
		\centering
		\includegraphics[width= 0.8\textwidth]{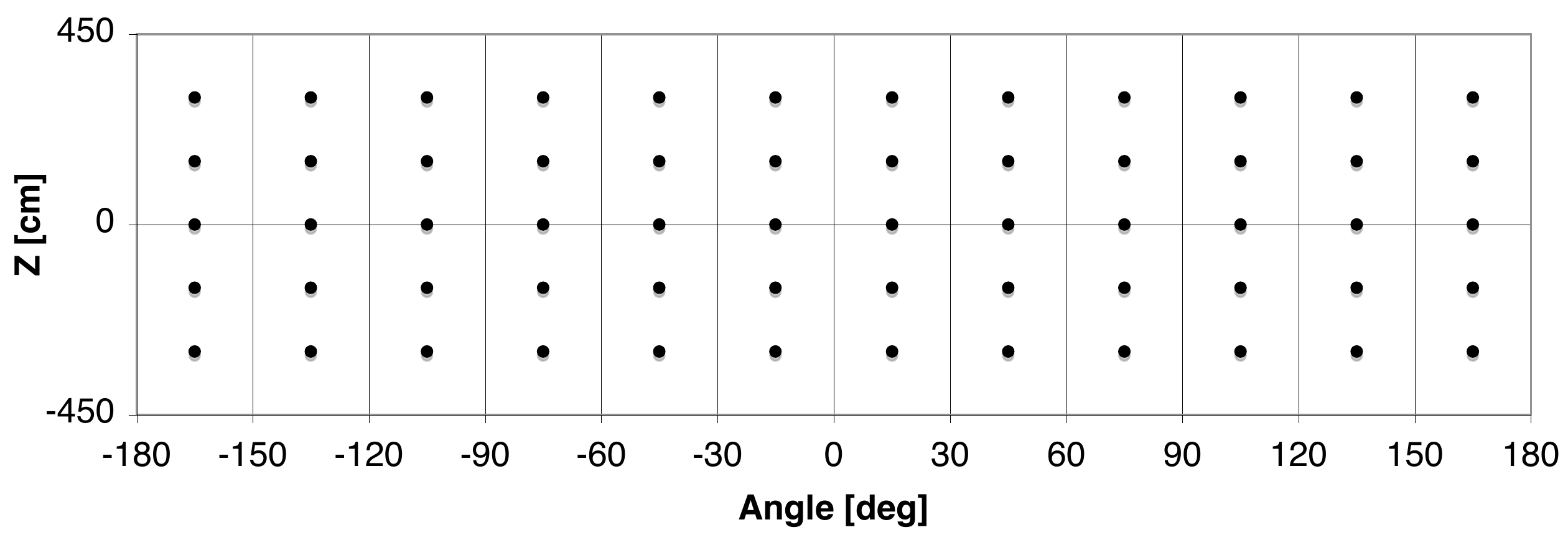}\\
		\footnotesize{\textsf{\textbf{f.} Lateral Array Model 1.}}
	\end{minipage}\\
	\vspace{0.5cm}
	\begin{minipage}[t]{1.0\linewidth} 
		\centering
		\includegraphics[width= 0.8\textwidth]{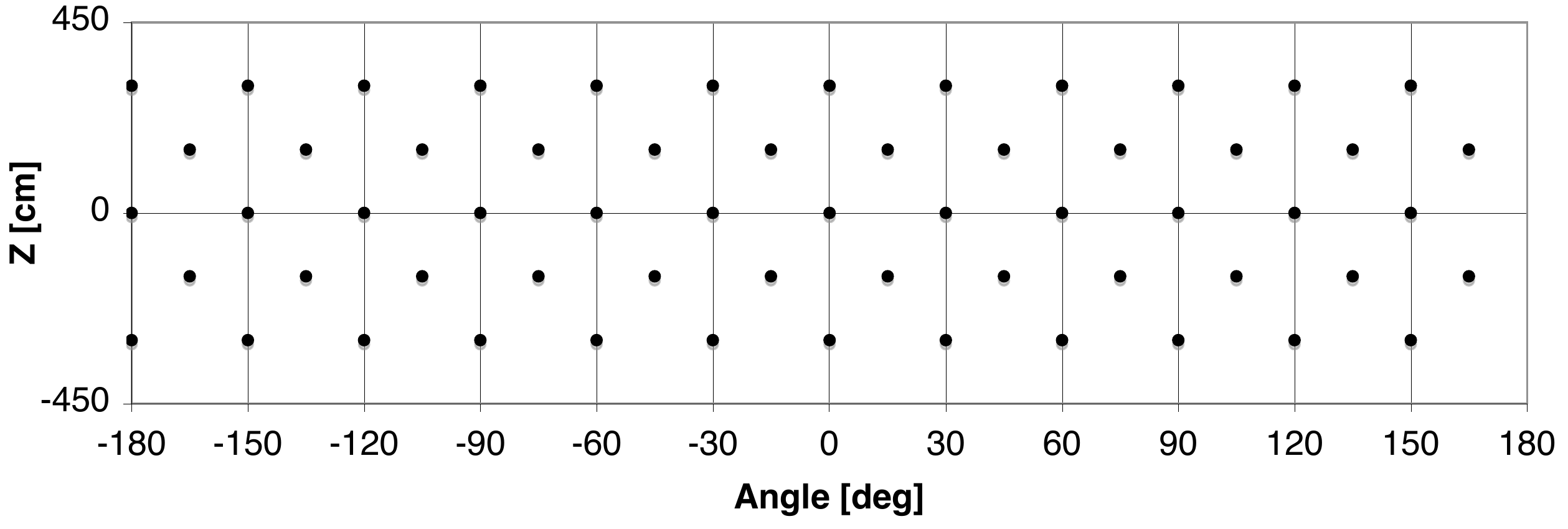}\\
		\footnotesize{\textsf{\textbf{g.} Lateral Array Model 2.}}
	\end{minipage}\\
	\caption{All the different PMT arrays tested.}
	\label{fig:PMTsGrids}
\end{figure}

Hit~patterns of the Cherenkov photons on the internal surface of the WT were generated for different configurations of reflector type and reflectivity levels. Then, with fixed reflector type and reflectivity, we evaluated the efficiency of the muon veto, defined as the percentage of tagged events over all the events hitting the WT, varying the PMT geometry arrangement. The geometries tested are displayed in figure~\ref{fig:PMTsGrids}.  After that, with the optimized PMT placement we checked the variation of the efficiency by varying the number of PMTs employed. All these models were compared with 80 different trigger conditions. The trigger conditions were requiring from 2 to 5 PMTs in coincidence to trigger an event, and, in each case, from 1 to 20 photoelectrons (PE) to trigger a PMT. The results obtained are presented in the next paragraph, the errors quoted are the statistical errors.

\vspace{0.25cm}

\section{Simulation results} \label{sec:results}

\subsection{Reflector type and reflectivity values} \label{sec:results:reflector}
The study of the hit~pattern of the photons on the internal surface of the WT shows that the diffusive reflector possesses a systematically lower photon survival percentage and a lower vetoing power efficiency compared to the specular one; this led to the choice of the specular reflector DF2000MA.

Some of the simulations results are presented in tables ~\ref{subtable:light_mu} and \ref{subtable:light_shower}  in terms of illumination density (in photons~/~muon~/~m$^{\mathrm{2}}$) and veto efficiency \footnote{The efficiency values have been calculated assuming 84 PMTs arranged with the optimized geometry that will be illustrated in paragraph \ref{sec:results:PMTs}} for different reflective foil models.

Removing completely the reflective foil significantly decreases the number of photons  (compare model~1 with model~5 in tables  \ref{subtable:light_mu} and \ref{subtable:light_shower}); this causes a drastic drop in the efficiency of the veto for the shower~event case. Smaller but still relevant efficiency loss ($\sim$~5\%  in the shower~event case) is seen when removing the reflective foil from the top surface (compare model~1 with model~3 on table \ref{subtable:light_shower}).
For this reason we decided to clad all the internal surfaces (top, lateral and bottom) of the WT with the reflective foil.

\begin{table}[!htb]
 \begin{subtable}{.5\linewidth}
      \centering
      \resizebox{6.5cm}{!}{ 
       \begin{tabular}{|c|cc|c|}
\hline
\multicolumn{4}{|c|}{MUON EVENT}\\
\hline
Model	&	\multicolumn{2}{c|}{Illumination density}	 	& 	Efficiency 	\\
 		&	\multicolumn{2}{c|}{[photons~/~muon~/~m$^{\mathrm{2}}$]}	 	& 	[\%]	\\
		&	Bottom 		&	 Lateral		&				\\ 
\hline
1		&	 2640 $\pm$ 40	& 1640 $\pm$ 20		&	99.80  $\pm$ 0.05\% \\
\rowcolor{grey}2		&	 2470 $\pm$ 30	& 1460 $\pm$ 20		&	99.78  $\pm$ 0.05\% \\
3		&	 2380 $\pm$ 30	& 1390 $\pm$ 20		&	99.75  $\pm$ 0.05\% \\
4		&	 2330 $\pm$ 30	& 1330 $\pm$ 20		&	99.71  $\pm$ 0.06\% \\
5		&	 1400 $\pm$ 20	& 655 $\pm$  9		&	96.3  $\pm$ 0.2\% \\
6		&	 2340 $\pm$ 40	& 1540 $\pm$ 30		&	99.61  $\pm$ 0.07\% \\	
\hline
\end{tabular}
}
     \caption{}
     \label{subtable:light_mu}
\end{subtable}
 \begin{subtable}{.5\linewidth}
      \centering
        \resizebox{6.cm}{!}{ 
        \begin{tabular}{|c|cc|c|}
\hline
\multicolumn{4}{|c|}{SHOWER EVENT}\\
\hline
Model 	&	\multicolumn{2}{c|}{Illumination density}	 	& 	Efficiency 	\\
 		&	\multicolumn{2}{c|}{[photons~/~muon~/~m$^{\mathrm{2}}$]}	 	& 	[\%]	\\
		&	Bottom 		&	 Lateral		&				\\ 
\hline
1		&	 196 $\pm$ 12	& 203 $\pm$ 12		&	73.3  $\pm$ 0.5\% \\
\rowcolor{grey}2		&	 182 $\pm$ 11	& 169 $\pm$ 10		&	71.4  $\pm$ 0.5\% \\
3		&	 162 $\pm$ 10	& 162 $\pm$ 10		&	67.8  $\pm$ 0.5\% \\
4		&	 158 $\pm$ 10	& 143 $\pm$ 9		&	65.9  $\pm$ 0.5\% \\
5		&	 96 $\pm$ 8		& 67 $\pm$  4		&	42.0  $\pm$ 0.5\% \\
6		&	 177 $\pm$ 11	& 144 $\pm$ 8		&	67.2  $\pm$ 0.5\% \\	
\hline
\end{tabular}
}
\caption{}
\label{subtable:light_shower}
\end{subtable}
    \caption{Bottom and lateral illuminations and efficiency$^\textrm{\scriptsize{1}}$ for the muon event (a) and the shower event (b), in the following reflective foil models: 
\mbox{\hspace{1.85cm}1: Specular reflector. Reflectivity: 100\% top - 100\% bottom and lateral} \mbox{\hspace{1.85cm}2: Specular reflector. Reflectivity: 90\% top - 95\% bottom and lateral}
\mbox{\hspace{1.85cm}3: Specular reflector. Reflectivity: 0\% top - 100\% bottom and lateral}
\mbox{\hspace{1.85cm}4: Specular reflector. Reflectivity: 0\% top - 95\% bottom and lateral}
\mbox{\hspace{1.85cm}5: No reflector \qquad \qquad \qquad \qquad \qquad \qquad \qquad \qquad \qquad}
\mbox{\hspace{1.85cm}6: Diffusive reflector. Reflectivity: 90\% top - 95\% bottom and lateral}
\mbox{\hspace{0.0cm} 100\% reflectivity (ideal case) is presented for reference. The selected model is highlighted in grey.}
}
 \label{table:light_mu_shower}
\end{table}

The final efficiency calculation uses the reflectivity values reported in \cite{Chris} and shown in the curve in figure~\ref{reflectivity}.
In order to account for the difficulties in attaching the foil perfectly, especially on the top surface of the tank, we conservatively replace the values in the plateau of the curve between 400~nm and 600~nm, with the values of foil model~2.

\begin{figure}[H]
\centering
	\begin{minipage}[t]{0.45\linewidth} 
		\centering
		\includegraphics[width= 0.95\textwidth]{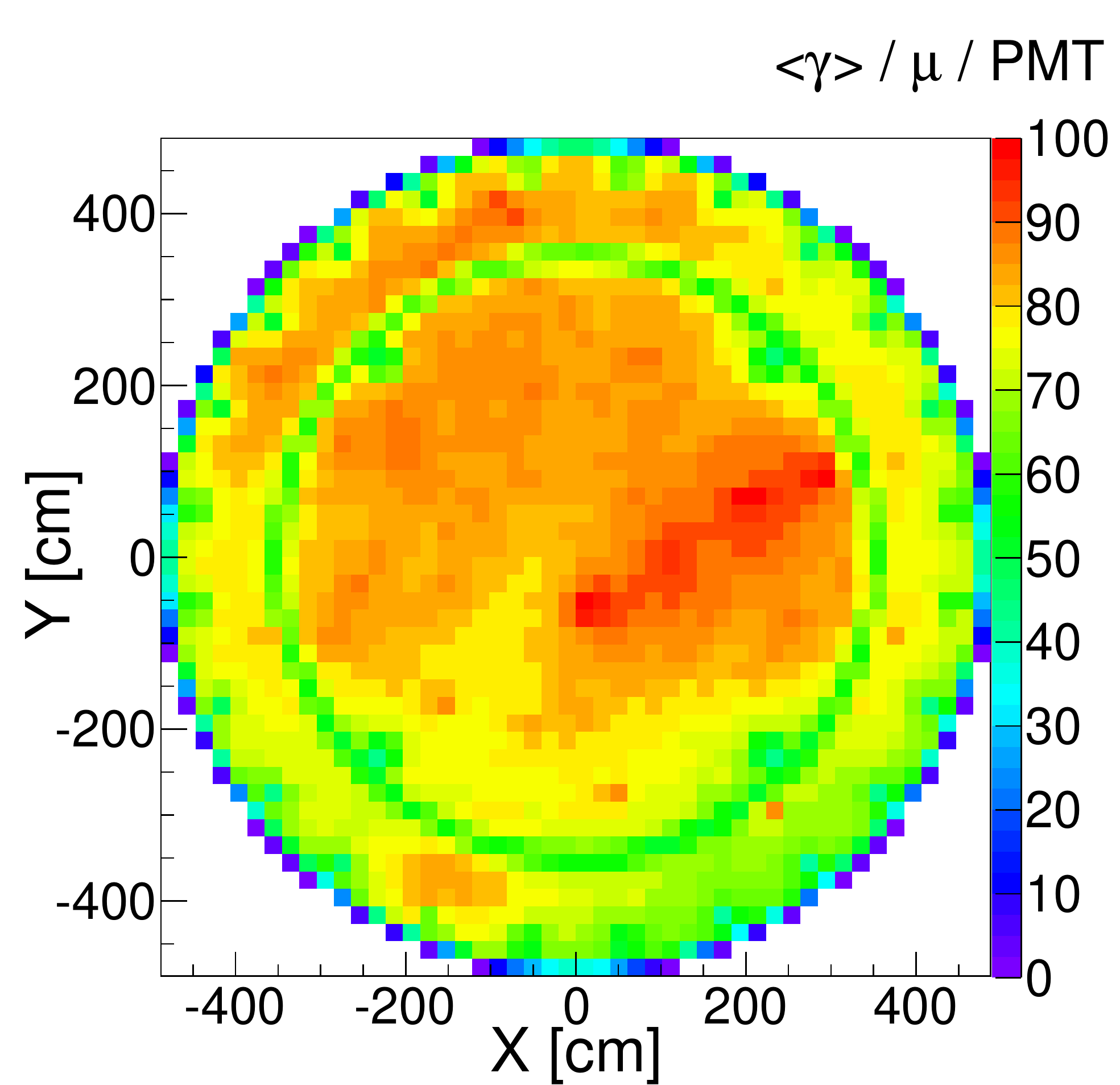}\\
		\footnotesize{(a)}
		\end{minipage}
		\hspace{0.5cm} 
		\begin{minipage}[t]{0.45\linewidth} 
		\centering
		\includegraphics[width= 0.95\textwidth]{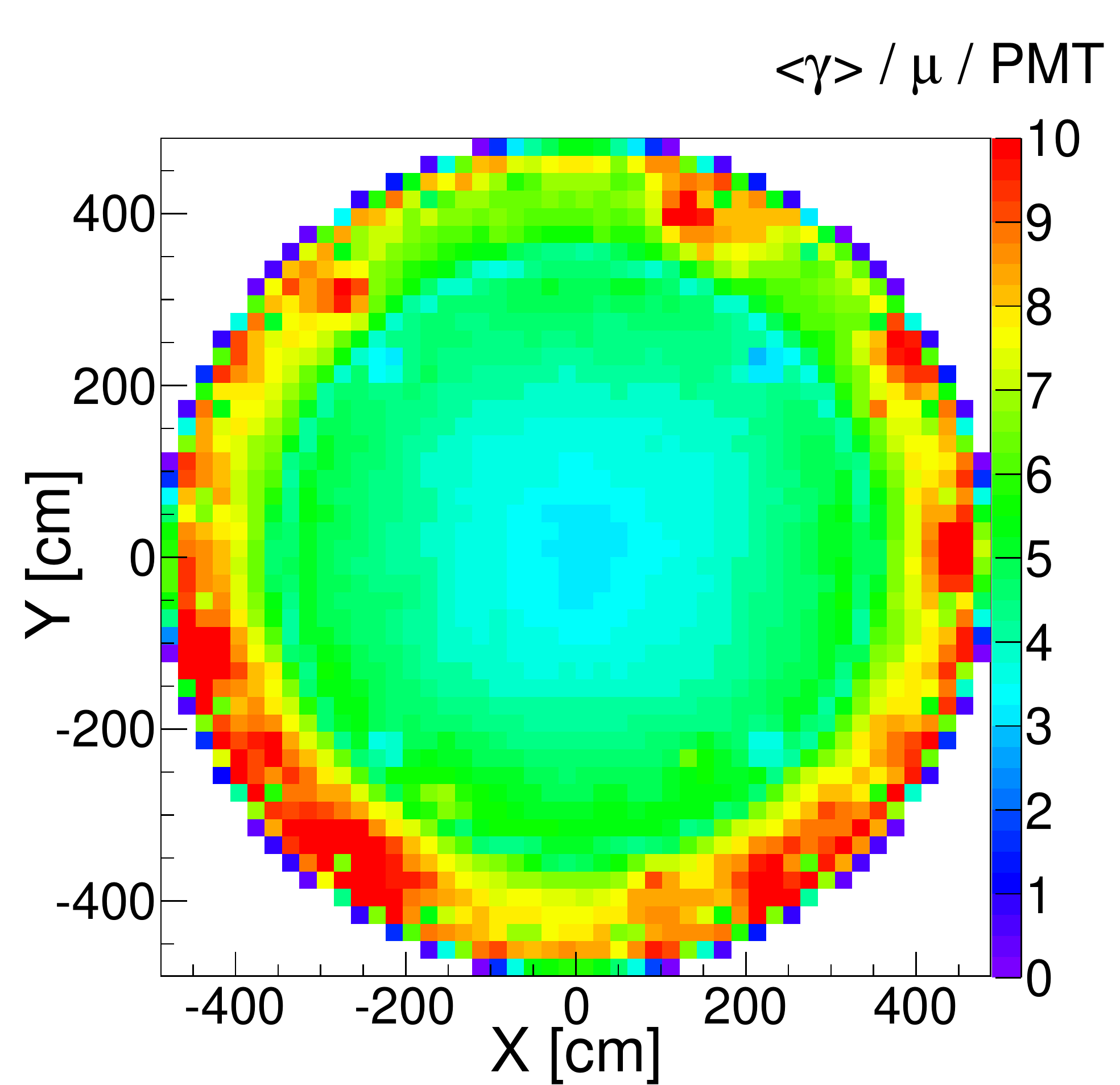}\\
		\footnotesize{(b)}
		\end{minipage}\\
	\vspace{1.0cm}
	\begin{minipage}[t]{1.0\linewidth} 
		\centering
		\includegraphics[width= 0.95\textwidth]{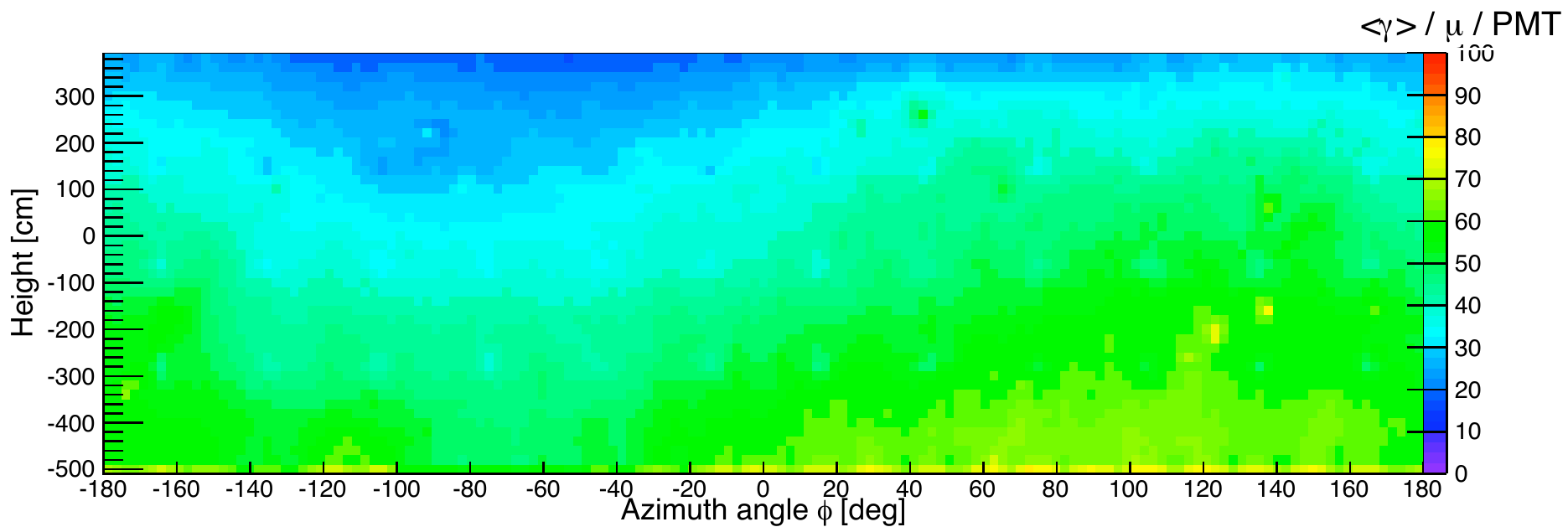}\\
		\footnotesize{(c)}
		\end{minipage}
		\hspace{0.5cm} 
		\begin{minipage}[t]{1.0\linewidth} 
		\centering
		\includegraphics[width= 0.95\textwidth]{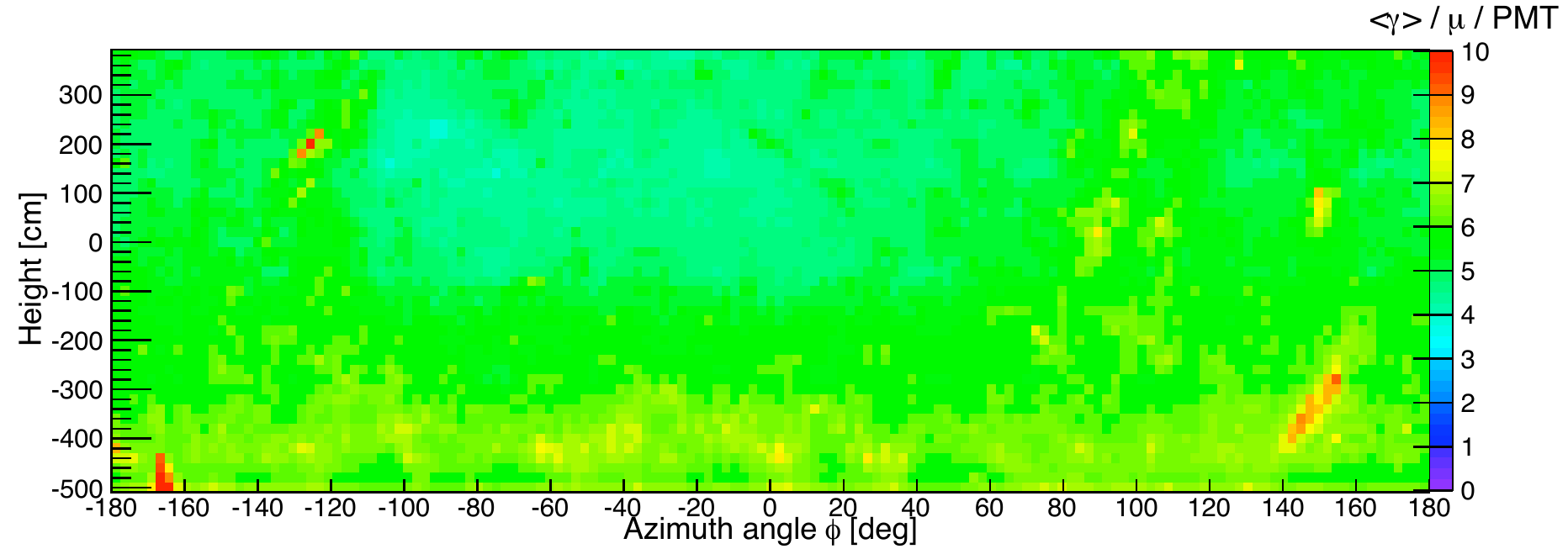}\\
		\footnotesize{(d)}
		\end{minipage}\\	
	\caption{Bottom and lateral illuminations for muon events, (a) and (c) respectively,  and shower events, (b) and (d) respectively, in the selected model (n.2 in table~\protect\ref{table:light_mu_shower}). Illumination values averaged over $10^{4}$ primary events.} 
	\label{fig:Illumination_mu}
	\end{figure}

In figure \ref{fig:Illumination_mu}, the distributions of Cherenkov photons for the bottom, (a) and (b), and the wall, (c) and (d), of the veto tank are shown. For each location, muon and shower types of events are displayed. The results are averaged over 10${^4}$ events.

The shadow of the objects very close to the surface of the tank can be seen in the distributions.  For example, the circular pipe of radius 300~cm of the water recirculation, located on the bottom surface of the tank is clearly visible in figures (a) and (b). Also the four feet of the support structure can be seen in figures (a) and (b) at positions always at $\pm 200$\,cm in X and Y. The difference in the light distribution between figures (a) and (b) is due to the fact that muons generally do not get absorbed in the WT. They produce light via Cherenkov effect all along their path. On the contrary, the shower particles get absorbed (or go below Cherenkov threshold) shortly after they travel in the water, and therefore, their Cherenkov light production takes place close to the entrance position in the tank. In both the bottom (a) and the lateral (c) illumination distributions for muon events, the effect of the muon directionality due to the profile of the mountain rock can be observed. The bright traces in figure (d) are due to shower events containing pions, decaying into muons while crossing the WT. These muons can produce intense Cherenkov light-traces, sometimes very spatially concentrated. This effect can alter the illumination density in the shower distribution even after mediating over 10$^{4}$ events.


\subsection{Arrangement and total number of PMTs} \label{sec:results:PMTs}
\begin{figure}[!b] 
  \centering 
   \includegraphics[height=7.cm]{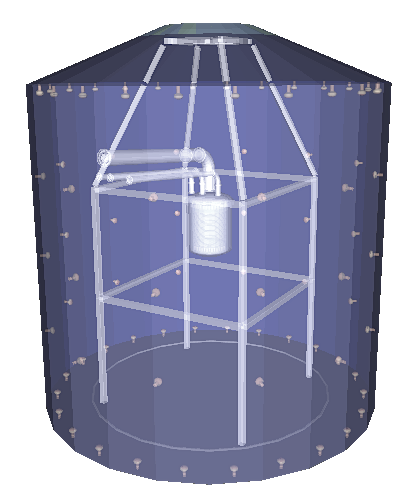} 
  \caption {The XENON1T geometry as used in the GEANT4 Monte Carlo simulation for this study.} 
  \label{fig:XENON1T_ MuonVetoMC3D} 
\end{figure}

The comparison of all the tested PMT geometries led to the conclusion that the best way to arrange PMTs, both for the muon~event case and for the shower~event case, is (see figure \ref{fig:XENON1T_ MuonVetoMC3D}):
\begin{itemize}
	\item TOP ARRAY: a ring at vertical height of 9~m from ground floor (the top edge of the cylindrical part of the tank), at radius 4.5~m, with the PMTs looking downward;
	\item LATERAL ARRAY: rings equally spaced vertically, attached to the tank's surface, with the PMTs looking inward;
	\item BOTTOM ARRAY: a ring in the bottom surface of the tank, at radius 4.5~m, with the PMTs looking upward.
\end{itemize}

The results indicate that the bottom and top grids give the best contribution to the efficiency when the PMTs are arranged very close to the external border of the circular area, rather than when they are spread all over the area itself. The reasons are explained below:

\begin{itemize}
	\item  for the muons: the muons traveling $>$1~m in water are tagged with high efficiency due to the enormous amount of light they produce, and the only events that could escape tagging are so-called ``border events'' constituted by those muons that cross the WT just in the outermost layer of water. Placing PMTs in those border areas enhances the possibility to recover a fraction of those events.
	\item for the showers: most of the light from the shower events is deposited in the very external part of the water volume, so in that region we have the highest chance to capture these more elusive events.
\end{itemize}

The results show that no difference exists in placing the lateral grids vertically aligned (like in figure~\ref{fig:PMTsGrids}f) or staggered (like in figure~\ref{fig:PMTsGrids}g). For mechanical reasons, it is easier to construct rails for the PMT holders in vertical alignment, so this was the one adopted.
Moreover, no improvement was obtained using a 45-deg inclination upward compared to the case with the PMTs looking inward.

With the above optimized arrays, a different total number of PMTs were tried.
In particular, high performances, (99.61$\pm$0.07)\%, in tagging ``muon~events'' were achieved with just 48 PMTs: 24 in the bottom array and 24 in the top array. 
Studying the ``shower~event'' tagging, we find that, in order to obtain good efficiency,  we need to increase the total number of PMTs by adding rings of PMTs to the lateral surface of the tank.

In table~\ref{table:tot_PMTs}, we present the results of the efficiency in tagging ``shower~events'' using 24 PMTs on the bottom and 24 PMTs on top, with the optimized arrangement just mentioned, and increasing the number of lateral PMTs from 0 to 3456, in steps of 12 PMTs per ring.
In the last entry of table \ref{table:tot_PMTs}, we have the configuration with the WT surfaces completely covered with PMTs. With this last configuration, we can estimate the fraction of `untaggable' events, that is, the fraction of events that, by their nature, produce so little light as to be invisible with whatever level of detector optimization. About~10\% of the shower~events are untaggable. Instead, to obtain at least 70\% efficiency in tagging showers, we need to employ 84 PMTs (24 in one ring in the top array, 36 in three rings of 12 PMTs each in the lateral array, 24 in one ring in the bottom array). This is the final configuration we decided to employ, as it is a reasonable compromise between performance and cost. 

\begin{table}[t!]
\centering
\resizebox{9.0cm}{!}{ 
\begin{tabular}{|c|c|c|c|c|}
\hline
\multicolumn{5}{|c|}{SHOWER EVENT}\\
\hline
Total  PMTs 	& Top array 	& Lateral array 		& Bottom array		& 	Efficiency [\%]	\\
\hline
48			& 24			& 0				& 24				&	 61.3 $\pm$ 0.5	 \\
60			& 24			& 12				& 24				&	 65.6 $\pm$ 0.5	 \\
72			& 24			& 24				& 24				&	 68.7 $\pm$ 0.5	 \\
\rowcolor{grey}84	& 24		& 36				& 24				&	 71.4 $\pm$ 0.5	 \\
96			& 24			& 48				& 24				&	72.6 $\pm$ 0.5	 \\
108			& 24			& 60				& 24				&	 74.7 $\pm$ 0.5	 \\
168			& 24			& 120			& 24				&	 79.5 $\pm$ 0.4	 \\
348			& 24			& 300			& 24				&	 85.3 $\pm$ 0.4	 \\
1848			& 24			& 1800			& 24				&	 89.0 $\pm$ 0.3	 \\
3504			& 24			& 3456			& 24				&	 89.9 $\pm$ 0.3	 \\
\hline \hline
9034			& 1589		& 5699			& 1746			&	 90.9 $\pm$ 0.3	 \\	
\hline
\end{tabular}
}
  \captionof{table}{Efficiency in tagging shower~events for different total number of PMTs (the grey box shows the selected case). The last values reference the configuration with a complete coverage of the WT with PMTs. The efficiency values were calculated assuming the triggering requirements of paragraph \protect\ref{sec:results:thresholds}.} 
   \label{table:tot_PMTs}
\end{table}

\begin{figure}[b!]
\centering
 \includegraphics[width=0.75\textwidth]{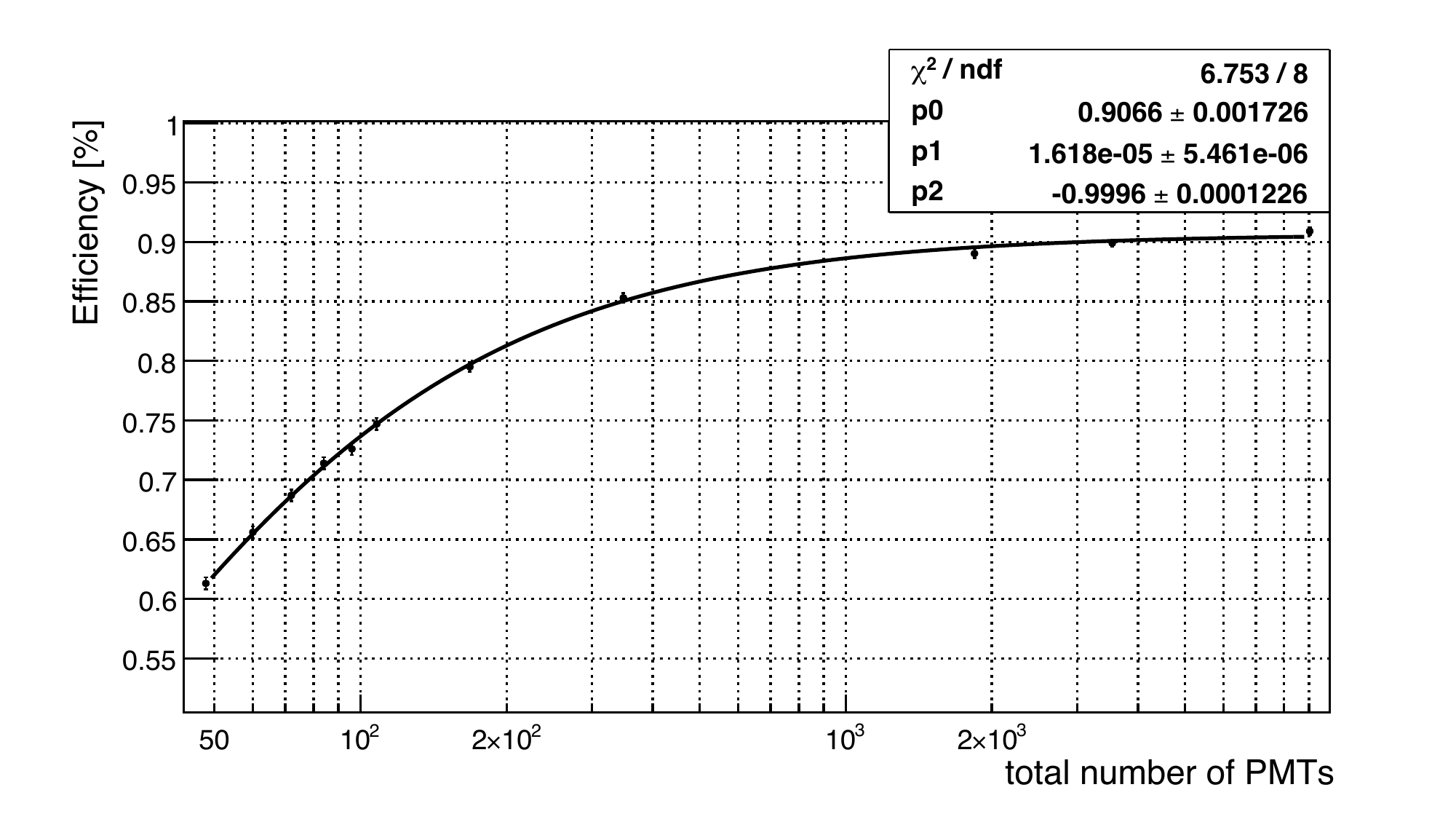}  
  \caption{Efficiency in tagging shower~events vs. an increasing number of PMTs employed in the previously optimized array geometry. The efficiency values were  calculated assuming the triggering requirements of paragraph \protect\ref{sec:results:thresholds}.} 
   \label{untaggable_fit}  
\end{figure}

The values in table~\ref{table:tot_PMTs} are plotted in the graph shown in figure~\ref{untaggable_fit} and fitted with a sigmoid function: \\
\begin{displaymath}
f(x) = p0 \cdot \frac{1-e^{-x \cdot p1}}{1+ p2 \cdot e^{-x \cdot p1}}
\end{displaymath}

\noindent where the parameters:
\begin{itemize}
	\item p0 = 0.9066 $\pm$ 0.0017
	\item p1 = (1.6 $\pm$ 0.5) $\cdot 10^{-5}$
	\item p2 = -0.99963 $\pm$ 0.00012
\end{itemize}

\noindent are related to:
\begin{itemize}
	\item p0: the limit of taggable events: $\sim$90\%  
	\item p1 and p2: 
	\begin{itemize}
		\item the ratio between the sensitive area (photocathode) and the hittable area (WT surface): $\sim1.9\cdot10^{-5}$;
		\item the impact of the reflective foil;
		\item the impact of the trigger configuration: 4 PMTs in coincidence within 300 ns (see paragraph \ref{sec:results:thresholds}).
	\end{itemize}
\end{itemize}

\vspace{2.0cm}

\subsection{Trigger condition}\label{sec:results:thresholds}
The signal from each PMT of the muon veto will be digitized at 100 MSamples/s. 
Two parameters determine the condition for which a single PMT participates in the trigger formation: 
\begin{itemize}
	\item threshold (\textit{thr}), the minimum amplitude that a single sample has to exceed. It is expressed as a fraction of the mean amplitude of a single PE; 
	\item time over threshold (\textit{tot}), the number of consecutive samples above threshold. 
\end{itemize}

\begin{figure}[!b]
	\centering
		 \vspace{0.5cm}
	\begin{tabular}{cc} 
	
		\hspace{-0.5cm} 
  		\begin{tabular}{c}
  			   \includegraphics[width= 0.44\textwidth]{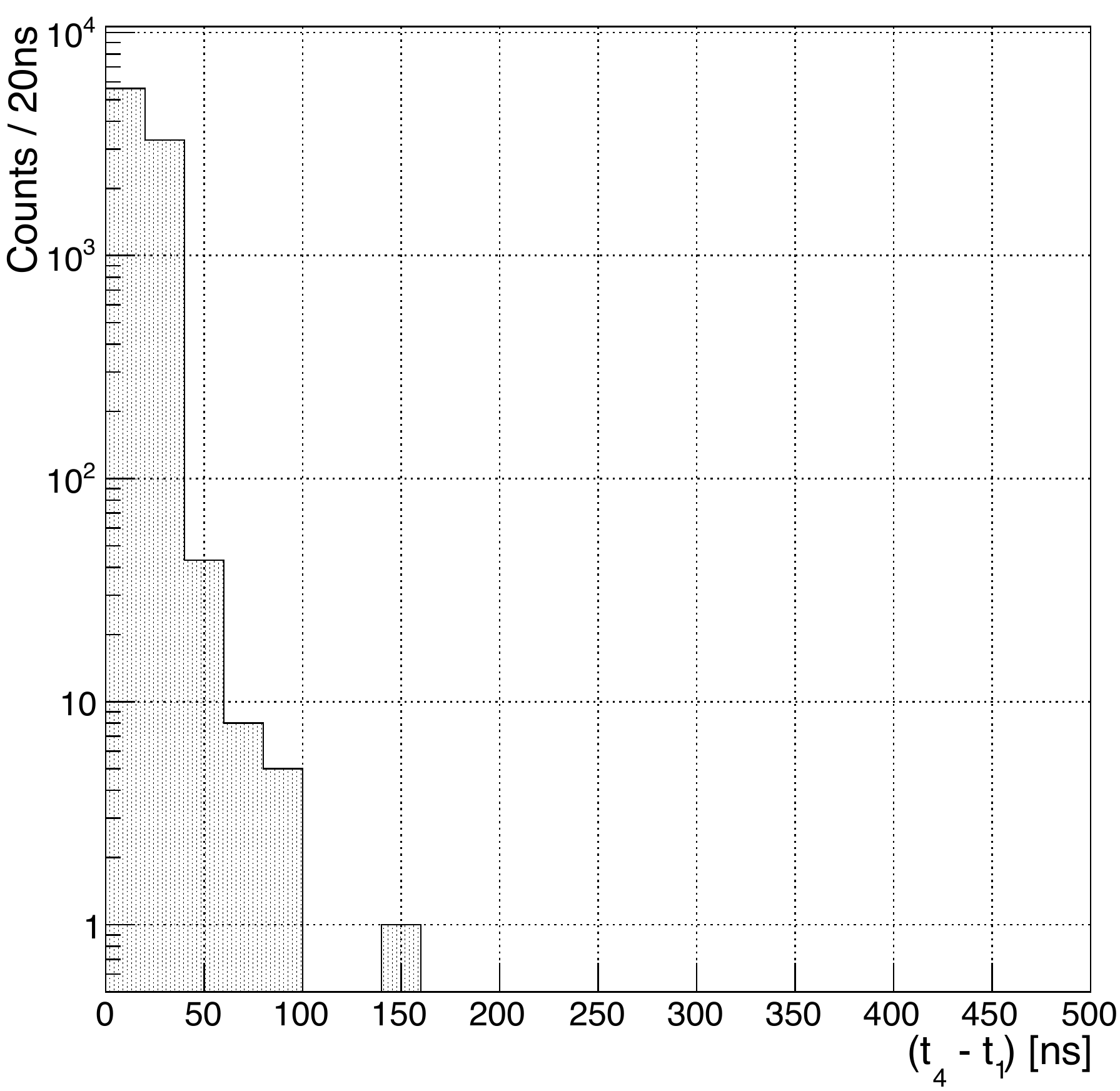} 
		\end{tabular}  

		&

  		\begin{tabular}{c}
  				\includegraphics[width= 0.44\textwidth]{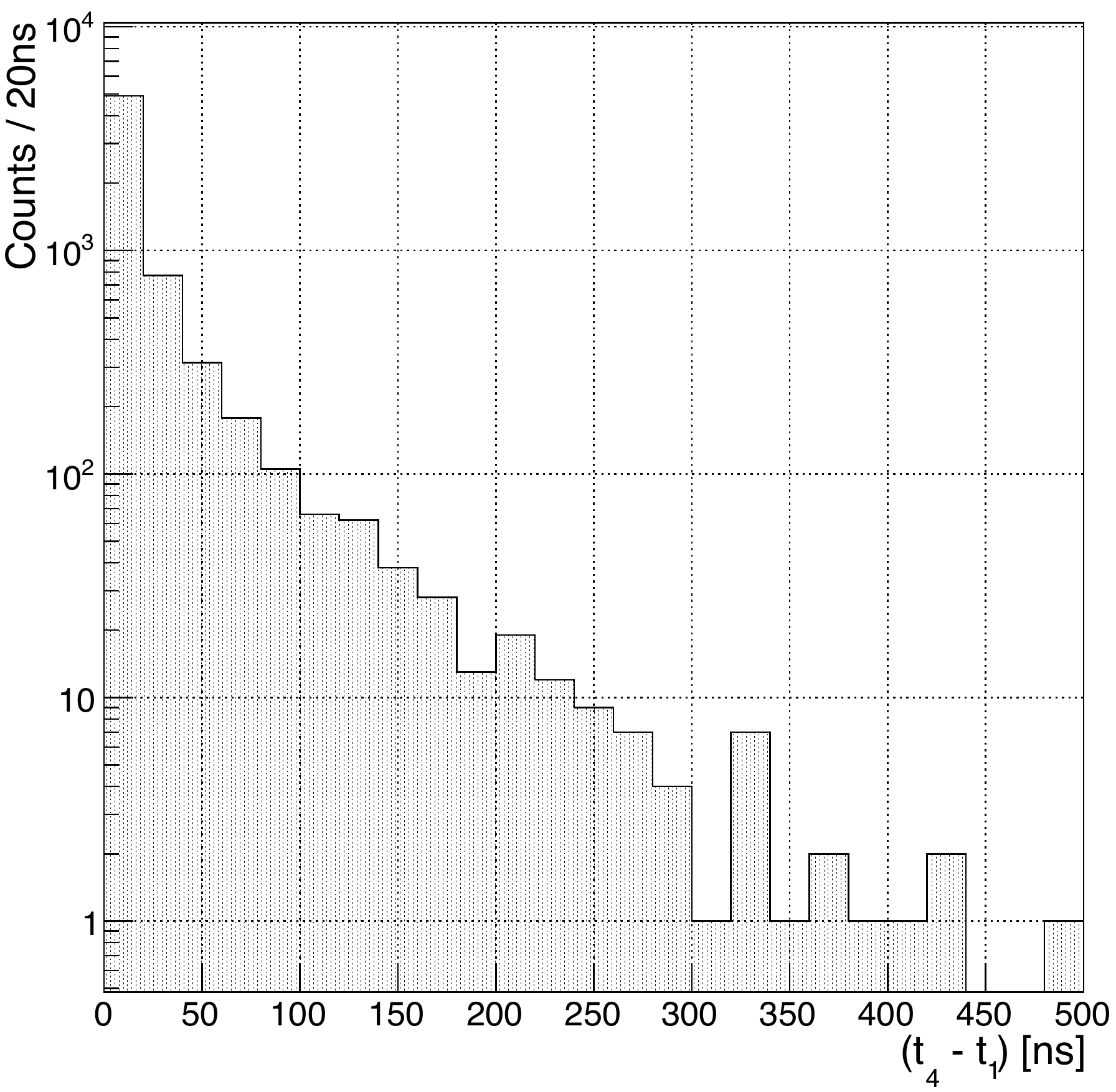} 
		\end{tabular}

		\\

		\scriptsize{(a)  Mu-event case:}

		 &

		\scriptsize{(b) Shower-event case}
		
		\\

	\end{tabular}
  \caption{Distribution of the maximum time difference of photons arriving on the PMT photocathodes in a 4-fold coincidence with the selected model (no.~2 of table~\protect\ref{table:light_mu_shower}) for the reflective foil and 84 PMTs in the optimized geometry.\\  
  	\mbox{$\bullet$ \hspace{0.1in} Primary events generated around the WT: $10^{4}$ in both cases.}\\
	\mbox{$\bullet$ \hspace{0.1in} Primary events hitting the WT: 8978 (a) and  9159 (b)}\\   
	\mbox{$\bullet$ \hspace{0.1in}  Fraction of the muon events hitting the WT, producing at least 4 PMTs on within 300~ns:} \\
	\mbox{\hspace{0.2in} 99.78\% (a) and 71.4\% (b).}
	} 
	   \label{Timing_84}
\end{figure}

A single PMT participates in forming the trigger if its signal has an amplitude greater than \textit{thr} for at least \textit{tot} consecutive samples. The trigger condition for the muon veto is the coincidence of $\ge$N PMTs inside a time window $\Delta$t, independently on PMT position inside the WT. The optimal width for $\Delta$t was determined by examining the distributions of the time difference of photons arriving on the PMT photocatodes in the muon and shower event simulations. An example is shown in figure \ref{Timing_84} for the 4-fold coincidence. In the following the time window is fixed to $\Delta$t = 300 ns.

\subsubsection{Tagging efficiency depending on trigger requirements}
\begin{figure}[!t]
	\centering
		
	\begin{tabular}{cc}

		\hspace{-0.5cm}
  		\begin{tabular}{c}
  			   \includegraphics[width= 0.44\textwidth]{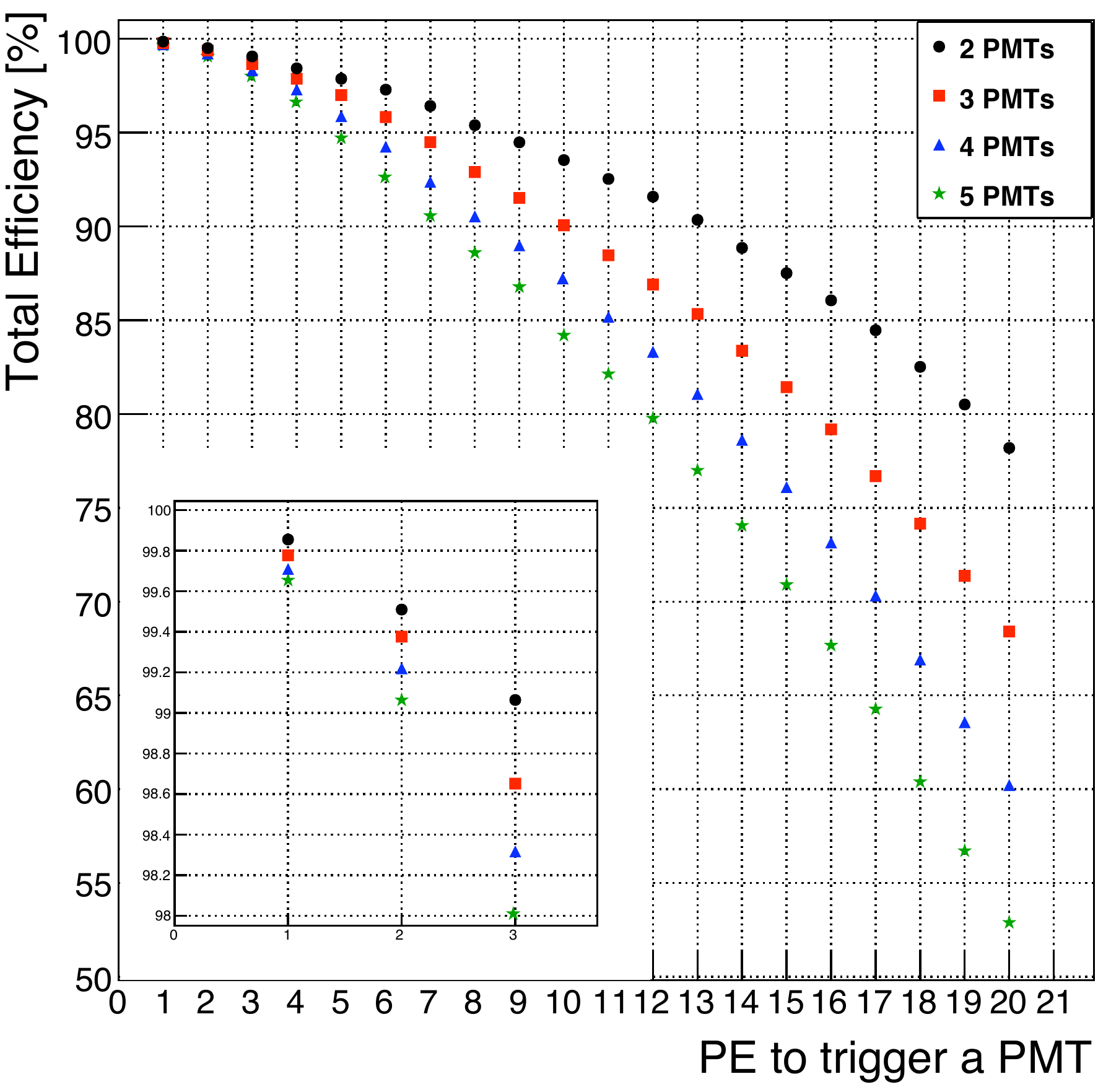}
		\end{tabular}  

		&

  		\begin{tabular}{c}
  				 \includegraphics[width= 0.44\textwidth]{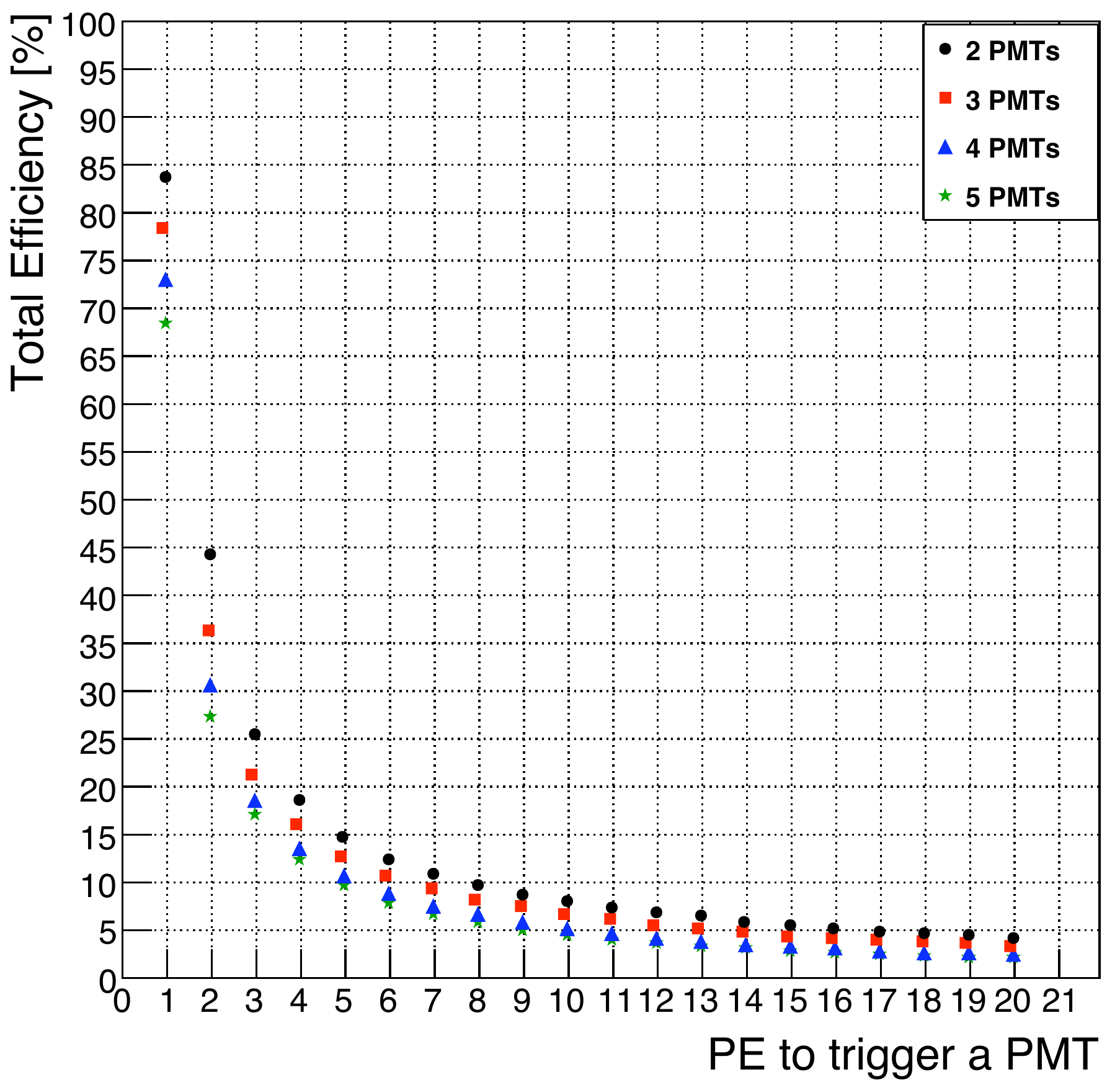} 
		\end{tabular}

		\\

		\scriptsize{(a) Muon-event case}

		 &

		\scriptsize{(b) Shower-event case}
		
		\\

	\end{tabular}
		\vspace{0.5cm}
  \caption{Curves of the efficiency versus the triggering requirements  with the selected model (no.~2 of table~\protect\ref{table:light_mu_shower}) for the reflective foil and 84 PMTs arranged with the optimized geometry.}
   \label{Efficiency_Triggering_84}
\end{figure}

The efficiency for tagging muon events and shower events depending on trigger requirements was obtained from Monte Carlo, with 84 PMTs arranged in the best configuration and with full efficiency in detecting single PE. 
Figure \ref{Efficiency_Triggering_84} shows the results depending on the number of PE detected by individual PMTs and for different number N of PMTs in coincidence. In order to keep a reasonable tagging efficiency for the shower events ($>$60\%), it is necessary to work at the single PE detection level. The choice of the number N of PMTs in coincidence is done with the goal to get the highest tagging efficiency while having a suitable trigger rate, as discussed in \ref{etrate}. Muon events are much less affected by the trigger conditions as their tagging efficiency is close to 100\% in many different cases.

\subsubsection{Expected trigger rate for the muon veto}\label{etrate}
\begin{figure}[t!]
	\centering
	\includegraphics[scale=0.6] {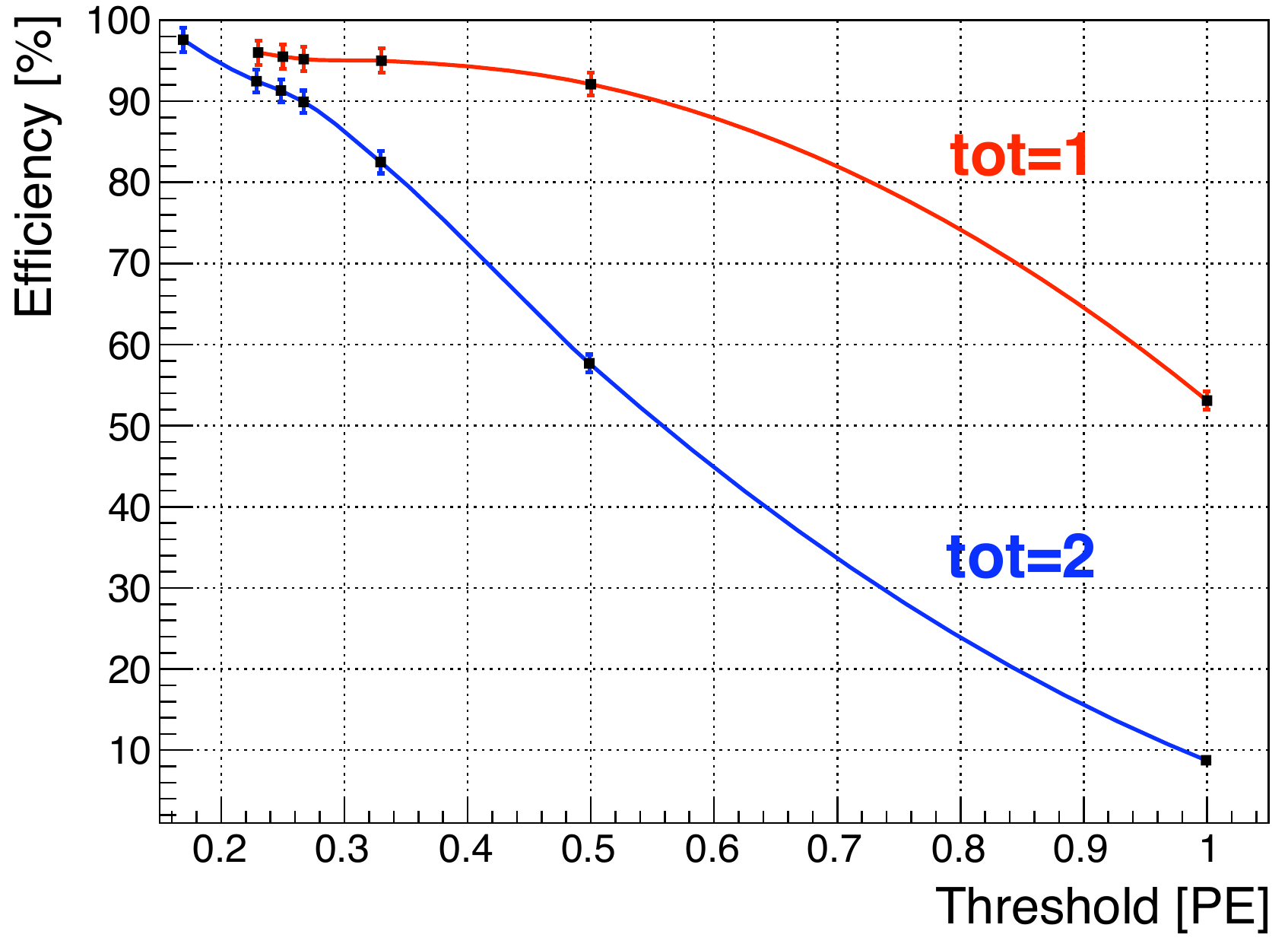}
	 \caption{The efficiency $\eta$ in detecting PE depending on  \textit{thr} for \textit{tot}~=~1 and \textit{tot}~=~2.}
	\label{fig:140403_efficiency}
\end{figure}

We studied the behaviour of the PMT dark count rate R$_{d}$ (\textit{thr}, \textit{tot}) with \textit{thr} varying in the interval 0.25-1.0 PE for \textit{tot}=1 and \textit{tot}=2.
$R_d$ varies between $R_d \sim 150$\,s$^{-1}$ ($thr$=1.0\,PE, $tot$=2) and $R_d~\sim~3000$\,s$^{-1}$ ($thr$=0.25\,PE, $tot$=1).
For each of these cases we determined the efficiency \textit{$\eta$} (\textit{thr}, \textit{tot}) for detecting a single PE, which is reported in figure \ref{fig:140403_efficiency}.
To get an estimate of the  counting rate of a single PMT in the muon veto, \textit{R$_{s}$}(\textit{thr}, \textit {tot}), we have to consider the contribution of the Cherenkov light induced by secondary electrons as a product of the interaction of gammas and neutrons from natural radioactivity in water. The flux of radioactive decay particles from rock and concrete in LNGS halls is assumed to be 1~$\gamma~/~\mathrm{cm^{2}~/~s}$ \cite{Arpesella, TesiSerena} and $5 \cdot 10^{-6}$~neutrons~/~cm$^{\mathrm{2}}$~/~s \cite{TesiSerena, Belli}. For the radioactive contamination of the stainless steel we use the results in \cite{Alfred}, conservatively multiplied by a factor 100. A dedicated Monte Carlo shows that the only important contribution comes from $\gamma$ emitted in the rock, which on average induce on each PMT a rate of R$_{b}\sim$2000 PE/s. 
Accidental coincidences occurring inside $\Delta$t are the largest contribution to the trigger rate for N$\le$5. These are calculated as follows:

\begin{displaymath}
	R_{\ge N} = \sum^{M}_{j=N} R_{j-1} \cdot (M -j +1) \cdot R_{s}(thr, tot) \cdot \frac{\Delta t}{(j-1)} 
\end{displaymath}
 
 \noindent where:

\begin{itemize}
	\item  $R_{s}(thr,tot)$ is the single PMT counting rate calculated as: $R_{s}(thr,tot) = R_{d} + \eta \cdot R_{b}$
	\item M=84 is the total number of PMTs
	\item R$_{1}$ = M $\cdot$ R$_{s}$.
\end{itemize}

We also expect to observe scintillation of $\alpha$ particles in the DF2000MA reflective foil. There are hints for the observation of this effect in the GERDA WT \cite{GERDA2}, where the same kind of reflective foil is installed. We derived an upper limit on the expected scintillation rate of $\alpha$ with dedicated measurements of the scintillating properties of the foil. The contribution to the muon veto trigger rate remains subdominant with respect to accidental coincidences for a number of PMTs in coincidence N$\le$5. 

\vspace{0.5cm}

Finally we have the trigger rate induced by muons passing through the WT (with or without accompanying neutron) and shower events. Considering the dimensions of the WT we may expect a rate R$_{\mu}\sim$5$\cdot$10$^{-2}$ s$^{-1}$ of muons passing through the WT and an even lower rate due to shower events.

\begin{figure}[t!]
	\centering
	 \includegraphics[scale=0.6] {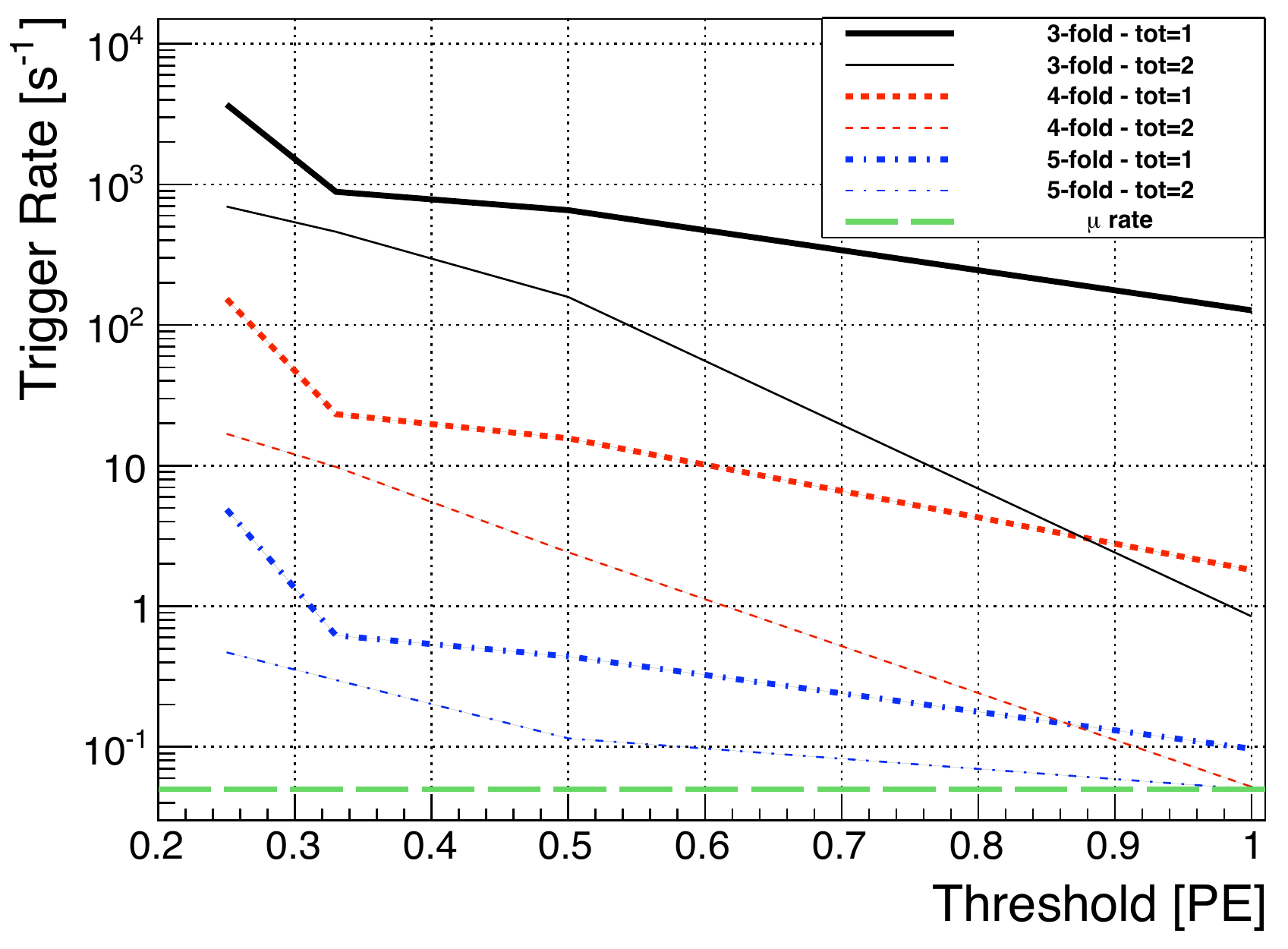} 
 	\caption{Muon veto trigger rate depending on \textit{thr}. The horizontal long dashed line marks the contribution from muon and shower events.}
	\label{fig:140403_MuonVetoTrigger_all_2}
\end{figure}

Figure \ref{fig:140403_MuonVetoTrigger_all_2} reports the muon veto trigger rate depending on $thr$ with the coincidence of $\ge$~N PMTs inside $\Delta$t. With N=3 the expected trigger rate is higher than 100~s$^{-1}$ at least when the efficiency for detecting PE is better than 50\%. With N=4 a good efficiency $\eta >$90\% can be achieved both with $tot$=1 and $tot$=2 while keeping the trigger rate lower than 20~s$^{-1}$. In the case N=5 the trigger rate can be lower than 1~s$^{-1}$ having high efficiency in detecting PE.

Muon veto and TPC will be synchronized by using a common digitizer clock.

\vspace{1.5cm}

\subsection{Summary of the final configuration} \label{sec:results:summary}
In conclusion we choose the following configuration:
\begin{itemize}
	\item Specular reflector (DF2000MA) with the reflectivity curve as measured in \cite{Chris} and shown in figure~\ref{reflectivity}, with the plateau of the curve between 400 and 600~nm modified according to model~2 of table~\ref{table:light_mu_shower};
	\item 84 PMTs (HQE 8''~Hamamatsu  R5912ASSY) with quantum efficiency, averaged over the Cherenkov light wavelength distribution, of about 30~\%  (figure~\ref{fig:QE_PMTs}), and arranged as in the conclusion of section \ref{sec:results:PMTs} (see figure \ref{fig:XENON1T_ MuonVetoMC3D});
	\item trigger at single PE (trigger efficiency 100\%) in a 4-fold coincidence of 300~ns time window.
\end{itemize}

\noindent The corresponding efficiency of the XENON1T water Cherenkov detector to veto muon-induced neutrons hitting the WT is:
\begin{itemize}
	\item (99.78 $\pm$ 0.05)\% for the ``muon~event'' case;
	\item (71.4 $\pm$ 0.5)\% for the ``shower~event'' case.
\end{itemize}

Considering these tagging efficiencies and the moderation and absorption of neutrons in the water shield, the residual neutron flux at the XENON1T cryostat surface is 1.2~$\cdot10^{-12}$~n/(cm$^{2}$~s). After applying general selection criteria for the WIMP search, namely that a valid event must yield a low-energy (5-50~keV) single scatter nuclear recoil interaction within an inner 1~ton fiducial liquid xenon (LXe) mass, we obtain a background of 0.01~events per year from muon-induced neutrons. This is in full agreement with the background goal of the XENON1T experiment, which is $<$1 background event in 2 ton$\cdot$year exposure, leading to an optimum sensitivity to spin-independent WIMP-nucleon scattering cross section of 2$\cdot$10$^{-47}$ cm$^{2}$ at m$_{\chi}$~=~50~GeV/c$^{\mathrm{2}}$ \cite{PaperoSelvi}.

\subsection{XENON1T upgrade: XENONnT}
After the main part of the study presented here was finalized, the collaboration decided to dimension most sub-systems of the XENON1T experiment such that it can easily be upgraded to an instrument featuring more than 7~tonnes of LXe. In particular all LXe handling systems, the detector support structure, as well as the outer cryostat are now already larger than required for the initial phase. We have verified that the results for the tagging efficiencies presented here are unaffected by these changes: the muon-tagging efficiency is identical for XENON1T and the upgraded detector. In case of showers, the efficiency slightly reduces from (71.4~$\pm$~0.5)\% to (70.6~$\pm$~0.5)\%.

\vspace{-0.20cm}

\section{Conclusions} \label{sec:end}
For the future XENON1T detector, the sensitivity goal of $2 \cdot 10^{-47} ~ \mathrm{cm^{2}}$ implies a  background reduction of two orders of magnitude compared to the level of the current XENON100 detector. This background reduction requirement, together with its location at the LNGS laboratory, necessitates the construction of  a muon veto. 

The type of muon veto chosen is a water Cherenkov detector based on a WT equipped with PMTs and a reflector foil. We have studied the optimization of its efficiency in tagging muon-induced neutrons through the Cherenkov light produced in water by the parent muon and/or the other muon-induced secondary particles. Specifically, we determined the most cost effective reflector type (studying also different reflectivity levels), PMT number and arrangement, and signal thresholds, to optimize the tagging efficiency, in both cases where we have or do not have the muon accompanying the neutron in the tank. The best choice for the reflector is a specular one and, for the PMTs, a good choice is a total of 84 arranged in 5 grids (1 top, 3 lateral and 1 bottom).

We found that it is feasible to work at the single PE detection level on a single PMT by requiring a 4-fold or 5-fold coincidence in a 300~ns time window.

The efficiency obtained in tagging muons that enter the muon veto detector is $>99.5\%$; these events represent $\sim$~1/3 of the cases where we have the muon-induced neutron inside the WT. For the other $\sim$~2/3 of the cases, thanks to the Cherenkov light produced in the water by secondary particles of the muon-induced shower, we are able to reach a tagging efficiency of  $>70\%$. The results remain valid for the upgrade of the experiment, XENONnT, with a target mass of up to 7~tonnes.
Combining the tagging efficiencies derived here with a full simulation of the propagation of muon-induced neutrons through the water tank and the LXe detector, we can predict a background of 0.01~event/year in the WIMP search region from this channel. Compared to the the background goal of $<$1 event for an exposure of 2 tons$\cdot$year, required to reach the design sensitivity, this is completely negligible.
None of these results take into account the wavelength shifting power of the reflector foil and are based on reflectivity values lower than that declared by the foil manufacturer.



\vspace{-0.20cm}

\acknowledgments
We gratefully acknowledge support from NSF, DOE, SNF,  FCT, Region des Pays de la Loire, STCSM, NSFC, BMBF, MPG, Stichting voor Fundamenteel Onderzoek der Materie (FOM), the Weizmann Institute of Science, I-CORE, the EMG research center and INFN. We are grateful to LNGS for hosting and supporting XENON1T. 
Moreover S.F. wishes to thank A.~Manzur, L.~Pandola and A.~Fanelli for advices and suggestions during the realization of this study.

\end{document}